\documentclass[fleqn,usenatbib]{mnras}

\usepackage{newtxtext,newtxmath}

\usepackage[T1]{fontenc}

\DeclareRobustCommand{\VAN}[3]{#2}
\let\VANthebibliography\thebibliography
\def\thebibliography{\DeclareRobustCommand{\VAN}[3]{##3}\VANthebibliography}

\usepackage{graphicx,times}
\usepackage{natbib}
\usepackage{amsmath}
\usepackage{multirow}
\usepackage{xcolor}
\bibpunct{(}{)}{;}{a}{}{,}

\title[Multiband decomposition of M\,51 with spiral arms]{Galaxies decomposition with spiral arms -- II: A multiwavelength case study of M\,51}

\author[A. A. Marchuk et al.]{Alexander A. Marchuk,$^{1,2}$\thanks{E-mail: a.a.marchuk@gmail.com}
Ilia V. Chugunov,$^{1,2}$
George A. Gontcharov,$^{1}$
Aleksandr V. Mosenkov,$^{3}$
\newauthor
Vladimir B. Il’in,$^{1,2,4}$
Sergey S. Savchenko,$^{1,2,5}$
Anton A. Smirnov,$^{1,2}$
Denis M. Poliakov,$^{1,2}$
\newauthor
Jonah Seguine,$^{3}$
Maxim I. Chazov$^{2}$
\\
$^{1}$Central (Pulkovo) Astronomical Observatory, Russian Academy of Sciences, Pulkovskoye chaussee 65/1, St. Petersburg 196140, Russia\\
$^{2}$St.Petersburg State University, 7/9 Universitetskaya nab., St.Petersburg, 199034, Russia\\
$^{3}$Department of Physics and Astronomy, N283 ESC, Brigham Young University, Provo, UT 84602, USA\\
$^{4}$Saint Petersburg University of Aerospace Instrumentation, Bol. Morskaya ul. 67A, St. Petersburg 190000, Russia\\
$^{5}$Special Astrophysical Observatory, Russian Academy of Sciences, 369167 Nizhnĳ Arkhyz, Russia\\
}

\date{Accepted XXX. Received YYY; in original form ZZZ}

\pubyear{2023}

\begin{document}
\label{firstpage}
\pagerange{\pageref{firstpage}--\pageref{lastpage}}
\maketitle

\begin{abstract}
Spiral structure can contribute significantly to a galaxy's luminosity. However, only rarely are proper photometric models of spiral arms used in decompositions. As we show in the previous work, including the spirals as a separate component in a photometric model of a galaxy would both allow to obtain their structural parameters, and reduce the systematic errors in estimating the parameters of other components. Doing so in different wavebands, one can explore how their properties vary with the wavelength. In this paper, second in this series, we perform decomposition of M\,51 in 17 bands, from the far UV to far IR, using imaging from the DustPedia project. We use the same 2D photometric model of spiral structure where each arm is modelled independently. The complex and asymmetric spiral structure in M\,51 is reproduced relatively well with our model. We analyze the differences between models with and without spiral arms, and investigate how the fit parameters change with wavelength. In particular, we find that the spiral arms demonstrate the largest width in the optical, whereas their contribution to the galaxy luminosity is most significant in the UV. The disk central intensity drops by a factor of 1.25--3 and its exponential scale changes by 5--10\% when spiral arms are included, depending on wavelength. Taking into account the full light distribution across the arms, we do not observe the signs of a long-lived density wave in the spiral pattern of M\,51 as a whole.
\end{abstract}

\begin{keywords}
galaxies: fundamental parameters -- galaxies: individual: M\,51 -- galaxies: spiral -- galaxies: structure
\end{keywords}

%
\section{Introduction}           
\label{sect:intro}

Galaxies are composed of many individual subsystems, which can each have distinct kinematics, spatial properties, stellar population ages, metallicities, and other properties. These observed differences are believed to be driven by various mechanisms which govern how galaxies form and evolve. By increasing our understanding of these subsystems, one can shed light on the role of major and minor merging, cold gas accretion, angular momentum transfer, bar formation, gravitational instabilities, and other local and secular processes \citep{2015ARA&A..53...51S}. 
\par
Decomposition is the process of highlighting, separating, and measuring parameters of galactic subsystems. It is usually done by modeling a galaxy with some functions and fitting the function parameters to best agree with models and observations. For example, one can use a 1D photometric cut, an azimuthally-averaged profile, or a 2D distribution of signal on image \citep{2015ApJ...799..226E,2017A&A...598A..32M}. 
The two most prominent subsystems in spiral galaxies are the central concentration component, called the bulge, and a large mostly flat component, called the galactic disk. Usually, a de Vaucouleurs or more general S{\'e}rsic profile is adopted for the bulge, and an exponential profile is adopted for the disk. The bulge-disk type of decomposition is the most common and has been carried out for the largest samples of galaxies \citep{2011ApJS..196...11S,2016ascl.soft04008L,2014ApJ...787...24B,2022MNRAS.511.3063M}. Such simple profiles allow one to do the decomposition conveniently and identify many physical relations, such as the connection between supermassive black hole and bulge properties \citep{2012MNRAS.419.2264V}, the dependence of the shape of the bulge intensity distribution on its luminosity \citep{2007MNRAS.376.1480N}, the dependence of the bulge and disk properties on their environment (i.e. whether they are situated in a field or in a cluster, \citealp{2009MNRAS.394.1991B}), galaxy evolution \citep{2009ApJ...696..411W}, and so on. 
\par
However, galaxies often contain additional components aside from the disk and bulge, and thus demonstrate a more complicated light distribution. Therefore, various studies adopt more sophisticated photometric models. For example, for a central component of a galaxy it is often necessary to add a second S{\'e}rsic profile \citep{2014MNRAS.443.1433D}. Sometimes, active galactic nuclei are modeled as an unresolved point source \citep{2008MNRAS.384..420G} or a nuclear disk \citep{2020A&A...643A..14G}. For edge-on galaxies, boxy or peanut-shaped (B/PS) bulges make frequent appearances \citep{2020MNRAS.499..462S,2022MNRAS.512.1371M} along with the barlens component in galaxies viewed more face-on \citep{2014MNRAS.444L..80L,2017A&A...598A..10L,2015MNRAS.454.3843A}. Disk components are usually modified by breaks of different types \citep{2014MNRAS.441.1992L}, or a truncation of the inner part which is justified from bar presence or quenching \citep{2022A&A...658A..74P}. Additional thick or thin disk components and flaring can also be added \citep{2021MNRAS.507.5246M}. 
\par
Despite the long list of aforementioned modifications, in some cases this too may not be enough to adequately represent a particular galaxy, since non-axisymmetric features may also be present \citep{2010AJ....139.2097P}. Bars and spirals are by far the two most common of these features, with more than half of all galaxies containing bars \citep{2011MNRAS.411.2026M,2008ApJ...675.1141S,2007ApJ...659.1176M} or spiral arms \citep{2006MNRAS.373.1389C,2013MNRAS.435.2835W}. Despite a similar universal abundance of such non-axisymmetrical features, bars are commonly accounted for in decomposition \citep{2009MNRAS.393.1531G}, whereas spirals are not. One reason for including bars but not spirals is the much higher relative contrast of bars, compared to typical spiral structures, and also the fact that bars extend through the central regions of galaxies thus their inclusion is crucial for obtaining the correct bulge parameters \citep{2009ApJ...696..411W,2015ApJS..219....4S,2009MNRAS.393.1531G,2007MNRAS.381..401L,2006AJ....132.2634L}. The other reason is that bars are relatively simple and could be modeled with a Ferrer's profile or a similar model \citep{2017MNRAS.469.4414W}, but spiral arms are much more difficult to accurately model.
\par
The spiral galaxies are usually divided into three classes following \citet{1990NYASA.596...40E}: grand design galaxies which contain two main spiral arms, multi-armed spirals with more than two distinct arms, and flocculent galaxies with many short and fuzzy arms. Arms can be symmetric and may demonstrate a good fit using a logarithmic curve, but often they do not follow such simplistic shapes \citep{2013MNRAS.436.1074S}. Spiral arms are probably among the largest visible features in galaxies, and can contain a large portion of the light budget, especially in grand design galaxies, where they can account for up to 40\% of the total luminosity \citep{savchenko}. Thus, it is natural to expect that such a bright component should affect the decomposition results, but 
it is not exactly clear how. This subject has been poorly investigated in the literature, and 
the results are quite controversial \citep{2022A&A...659A.141S,2014ApJ...780...69L,2017ApJ...845..114G}.
\par
Spiral arms' influence on decomposition mostly accounted for artificial images of galaxies \citep{2022A&A...659A.141S,2020ApJ...900..178L,2012ApJS..199...33D}. This approach is obviously flawed due of its simplicity and the unnatural appearance of modeled galaxies when compared to real objects. The only credible approach used for decomposition of observations with spiral arms that we are aware of is formulated in \citet{2010AJ....139.2097P}. Here, spirals are modelled as Fourier and bending modes, modified by a rotation function of different forms. Although this approach works surprisingly well even for some complex spiral galaxies, it is challenging to implement and interpret, and is thus only used in a few works with a very limited number of objects \citep{2014ApJ...780...69L,2017ApJ...845..114G}. Moreover, the rotation function is smooth, and, hence, not able to model observed pitch angle changes \citep{2013MNRAS.436.1074S}, where the pitch angle is the angle between the line tangent to the spiral arm and the line perpendicular to the radius-vector drawn from the centre of the galaxy. Also, the resultant spiral structure is symmetrical for each mode, which is not the case for many objects \citep{1997PASP..109.1251C}. Finally, the importance of spiral pattern accounting is not evident after these studies. 
\par
In our previous pivot study \citet{2024MNRAS.527.9605C}, first in the series, we attempt to address these issues by introducing a new 2D photometric model where each spiral arm is modeled independently. In this model, the light distribution both along and across the arm and its overall shape can be varied significantly. For 29 galaxies we utilize 3.6$\upmu$m-band images corresponding to the old stellar population from \cite{2015ApJS..219....4S}, and analyze the differences between models with and without spiral arms. We find that spiral arms are modeled well, and neglecting them in decomposition causes errors in estimating the parameters of the disk, the bulge, and the bar. In addition, \citet{2024MNRAS.527.9605C} find that the pitch angle of spiral arms decreases with increasing bulge or bar fraction, and that the spiral-to-total ratio is higher for galaxies with more luminous disks and with higher bulge-to-total ratios. In the first study we also measure the widths of the spiral arms and the contribution of the spiral arms to the azimuthally-averaged brightness profile. Overall, we demonstrate that the approach, presented in \citet{2024MNRAS.527.9605C}, produces reliable models and is useful for the analysis of spiral structure in observed galaxies.
\par
Decomposition is rarely done for more than a few bands. In most cases only adjacent bands from the same survey are used, such as SDSS optical bands \citep{2011ApJS..196...11S,2018MNRAS.473.4731K,2017A&A...598A..32M}, $JHK_s$ from 2MASS \citep{2021MNRAS.507.5952R}, or less frequently, {\it Herschel} far infrared 100-500~$\upmu$m bands \citep{2019A&A...622A.132M}. However, to model stellar populations of individual galaxy components and build their spectral energy distribution (SED), it is important to use the longest wavelength baseline. Using bands from different telescopes is difficult due to the additional processing required to account for different instrument resolution and other characteristics. For these reasons, studies with a wide wavelength range coverage from far ultraviolet (FUV) to far infrared (FIR) are scarce (see \citealp{2021MNRAS.504.2146B} and \citealp{2022MNRAS.513.2985R} as examples). There also exists an approach to do decomposition of images in different wavelengths simultaneously, i.e. to make single, wavelength-dependent model which fits images in all bands\citep{2014MNRAS.444.3603V,2013MNRAS.430..330H}, but so far it has only been applied to a limited number of bands.
\par
In this paper, we decompose images of the well-known Messier 51 (M\,51) galaxy taking into account not only bulge and disk, but also its spiral arms, using a modified analytical model from \citet{2024MNRAS.527.9605C} for this purpose. The images span a wavelength range from the FUV to FIR and encompass different sources of radiation, such as both young and old stars, and dust and polycyclic aromatic hydrocarbon (PAH) emission. Thus, we aim to (i) demonstrate that our model can fit a nuanced galaxy with non-symmetrical and complex arms, and (ii) do this for a wide set of photometric bands.
\par
The galaxy M\,51 (NGC\,5194, Whirlpool Galaxy) is an iconic and well-known object in astronomy as it was the first nebula with a detected spiral structure \citep{1850RSPT..140..499R}. The term M\,51 is often used to describe both NGC\,5194 (M\,51a) and its companion galaxy NGC\,5195 (M\,51b) as an interacting system, but we will use it as a shorthand for the main galaxy only. The system is a famous example of an Sbc galaxy with grand design arms, and a close lenticular companion NGC\,5195 is attached to the tip of the northern spiral arm. This apparent connection is actually a visual coincidence, since NGC\,5195 is located at 20-50~kpc behind M\,51 \citep{2010MNRAS.403..625D,2000MNRAS.319..377S,1972ApJ...178..623T}.  There is no full consensus about number and time of interactions between galaxies. Numerical models predict two interactions \citep{2000MNRAS.319..377S,2010MNRAS.403..625D} or a single flyby event about 100~Myr ago \citep{1972ApJ...178..623T}. In any case, the signs of recent interaction are clearly visible both in deep and H{\sc i} images as a tidal bridge that connects the galaxies, and as a large tail. Both galaxies host weak active galactic nuclei \citep{2001ApJ...560..139T}. The distance $D=8.9$ Mpc to M\,51 was derived using several SN{\sc ii} events \citep{2008ApJ...675..644D}, which gives the scale 43 pc/arcsec. We adopt an inclination angle $i=32.6\degr$ following \citet{radiativeM51}, who reuse the results presented in \citet{2019A&A...622A.132M}.
\par
We consider M\,51 in this study because it is close, so its various aspects are very well studied, and in as an homage to its historical importance in astrophysics. It was also an aim of decomposition processes in many previous works, namely \cite{2015ApJS..219....4S,2008AJ....136..773F,2010AJ....139.2097P,1998AJ....116.1626B,2017A&A...605A..18C,2022ApJ...930..170H}, which make it possible to compare the obtained parameters with previous estimations for models without spiral structure considered as separate components. Essential for this study is that M\,51 has a huge and contrasting grand design spiral pattern. Moreover, these spiral arms exhibit clear asymmetry and pitch angle change. Together with an overall complex morphology, this makes M\,51 an interesting and difficult target. 
\par
Our paper is organised as follows. In Sect.~\ref{sect:data} we  describe the galaxy and sample of images used. In
Sect.~\ref{sect:model} we present the model of spiral arm. In Sect.~\ref{sec:decomposition} we provide the details about decomposition and in Sect.~\ref{sec:validation} validate our models in different bands. In Sect.~\ref{sect:discussion} we present main results of our study along with the discussion. We summarize the work and give conclusions in Sect.~\ref{sec:conclusions}.

\section{Data}
\label{sect:data}

\begin{table}
\centering
\caption[]{Photometric bands used. Horizontal lines demarcate different parts of the spectrum, which are shown with filled color in Figures throughout the paper. FWHM and pixel scales are given in arcseconds. The last column marks if M\,51 was additionally decomposed in this band using an image with the original resolution.}
\label{tab:dustpedia_bands}
\begin{tabular}{l|c|c|c|c|c}
\hline\noalign{\smallskip}
Name/Facility & $\lambda$, $\upmu$m  & Pixel & FWHM & $\log_{10}(\lambda)$ & Orig. res.?\\
\hline
\hline\noalign{\smallskip}
GALEX FUV & 0.153 & 3.2 & 4.3 & -0.82 & + \\
GALEX NUV & 0.227 & 3.2 & 5.3 & -0.64 & + \\
\hline
SDSS $u$ & 0.353 & 0.45 & 1.3 & -0.45 & + \\
SDSS $g$ & 0.475 & 0.45 & 1.3 & -0.32 & + \\
SDSS $r$ & 0.622 & 0.45 & 1.3 & -0.21 & + \\
SDSS $i$ & 0.763 & 0.45 & 1.3 & -0.12 & + \\
SDSS $z$ & 0.905 & 0.45 & 1.3 & -0.04 & + \\
\hline
2MASS $J$ & 1.24 & 1 & 2.0 & 0.09  & -\\
2MASS $H$ & 1.66 & 1 & 2.0 & 0.22  & -\\
2MASS $K_s$ & 2.16 & 1 & 2.0 & 0.33  & -\\
\hline
Spitzer 3.6~$\upmu$m & 3.6 & 0.75 & 1.66 & 0.56 & + \\
Spitzer 4.5~$\upmu$m & 4.5 & 0.75 & 1.72 & 0.65 & - \\
\hline
Spitzer 8.0~$\upmu$m & 8.0 & 0.6 & 1.98 & 0.90 & + \\
Spitzer 24~$\upmu$m & 24 & 1.5 & 6 & 1.38 & + \\
Spitzer 70~$\upmu$m & 70 & 4 & 18 & 1.85 & + \\
PACS 160~$\upmu$m & 160 & 4 & 13 & 2.20 & + \\
SPIRE 250~$\upmu$m & 250 & 6 & 18 & 2.40 & + \\
\noalign{\smallskip}\hline
\end{tabular}
\end{table}

The source of images used in this study is the DustPedia  Archive\footnote{http://dustpedia.astro.noa.gr/}. Introduced in \citet{dustpedia},  DustPedia is a multiwavelength catalogue with photometry across 42 bands for the 875 galaxies in the local Universe. Based on objects observed by {\it Herschel} Space Observatory \citep{2010A&A...518L...1P}, it also includes images from the GALaxy Evolution eXplorer (GALEX, \citealp{2007ApJS..173..682M}), the Sloan Digital Sky Survey (SDSS, \citealp{2000AJ....120.1579Y}), the Digitized Sky Survey (DSS\footnote{https://archive.eso.org/dss/dss}), the 2 Micron All-Sky Survey (2MASS, \citealp{2006AJ....131.1163S}), the Wide-field Infrared Survey Explorer (WISE, \citealp{2010AJ....140.1868W}), the {\it Spitzer} Space Telescope \citep{2004ApJS..154....1W}, overall covering a huge part of the spectrum from FUV to FIR and submillimetre waves, spanning over five orders of magnitude in $\lambda$. All those observations were not only collected, but also processed for doing consistent mosaic co-addition, foreground star removal, aperture-matched photometry, noise reduction, unit conversion, and removing foreground emission. This enormous amount of work resulted in a convenient and effective tool to study multiwavelength imagery of galaxies without the burden of doing data reduction. Dustpedia is dedicated to study the dust role in extragalactic processes. It not only boosts that part of science \citep{2018A&A...620A.112B,2019A&A...622A.132M,2019A&A...623A...5D,2019A&A...624A..80N,2019MNRAS.489.5256C,2020MNRAS.494.2823T,2021MNRAS.506.3986N}, but could also be used for entirely different domains of astrophysics, such as galactic dynamics studies.
\par
We downloaded images for M\,51 from the DustPedia Archive in 17 different bands, listed in Table~\ref{tab:dustpedia_bands}. All images are in units of Jy/pix. We decided to not use DSS and WISE images due to their lower resolution and pixel width when compared to their close analogues, which are SDSS in the optical and Spitzer in the IR bands. To properly compare the properties of the spirals in the images obtained using different telescopes and to ensure that obtained insights are not introduced by differences between the instruments, we need to reduce all the images to the same resolution and point spread function (PSF).
Therefore, optimal resolution and depth are crucial for our analysis. We decide to not use bands beyond SPIRE 250~$\upmu$m, because near this wavelength, image resolution begins to deteriorate quickly. However, the 250~$\upmu$m band is located close to the peak of the SED in its FIR part where the emission from cold dust is not contaminated by the emission from warm dust, thus the inclusion of this band allows studying the properties of the cold dust in M\,51.
\par
Reducing all the images to the lowest resolution is methodologically correct, but has its price. The degradation of resolution could significantly affect the results, altering the smallest details and introducing a PSF-related bias \citep[e.g.][]{2021MNRAS.508.5825M}. To address this issue, we also perform decomposition of two additional samples of data. The first sample consists of images with the original resolution and PSF. These are listed in the last column of Table~\ref{tab:dustpedia_bands}. We do not use all bands to reduce the computational workload, but try to cover as wide $\lambda$ range as possible. The second sample consists of 14 images from the GALEX FUV to Spitzer 24~$\upmu$m, reduced to the same resolution. The choice of the 24~$\upmu$m limit is based on the fact that we deliberately do not want to omit the UV part of the spectra, but maximally improve the resolution. Since the largest pixel size is now for the GALEX data, this allows us to improve the resolution by a factor of two, and by a factor of three for the PSF FWHM. The inclusion of these two additional samples strengthens the results, presented throughout the paper, and provides a tool to catch the resolution-related effects as mentioned in Sect.~\ref{sec:bulge}. Note, however, that these datasets should be treated as auxiliary data only, because for them we do not estimate decomposition uncertainties using bootstrapping as described in Sec.~\ref{sec:decomposition} due to the computational burden of such a task.
\par
Most of the necessary image preparation routines have already been completed thanks to the DustPedia Collaboration, but we still need to  mention several of our methodological choices. Firstly, we note that the sky background in 2MASS images are very inhomogeneous, so we use data downloaded directly from 2MASS Atlas \citep{2006AJ....131.1163S} instead. These images' bands, $J, H, K_s$, were converted to Jy/pix and then used as is. Additionally, we note that some images in DustPedia demonstrate non-zero background emission, but it is still rather constant through the image and was thus subtracted. We do not de-project images into a face-on orientation using the same constant inclination value during decomposition, see Sec.~\ref{sec:decomposition} for details. We mask the foreground stars, the companion galaxy NGC\,5195, and the tidal tail where they were visible, manually and independently in each image.
\par
Secondly, we need to choose the band with the largest PSF full width at half maximum (FWHM), which is SPIRE 250~$\upmu$m in our case, and convolve the rest of the images with appropriate transition kernels such that after this convolution, the resulting images would have the same PSF FWHM. Thus, all images were resampled to the pixel width and FWHM of 250~$\upmu$m, which are 6~arcsec and 18~arcsec accordingly (for comparison, M 51's angular size in optical bands is about 10 arcmin). In order to estimate these transition kernels, we use information about PSFs published in \citet{2011PASP..123.1218A} and then apply the PSF matching tool from {\small PHOTUTILS} \citep{2020zndo...4044744B}. For the subsample limited by 24~$\upmu$m band we do exactly the same, but with the pixel scale and FWHM being 3.2~arcsec and 6~arcsec, respectively. The resulting PSF is very close to the PSF in the original 250 $\upmu$m image, including the PSF tail. In all cases, we resample the PSF to the same pixel size as in an image. We conducted some experiments with oversampled PSF and concluded that using an oversampled PSF has no influence on any results of decomposition.

\section{Analytical model}
\label{sect:model}

\begin{table}
\centering
\caption[]{Parameters in decomposition model. }
\label{tab:parameters}
\begin{tabular}{|l|c|c|}
\hline\noalign{\smallskip}
	Component & Parameter & Description \\
	\hline
    \hline\noalign{\smallskip}
	\multirow{3}{*}{All} & $X_0, Y_0$ & Coordinates of the galactic center \\
	 & PA & Position angle of galactic plane \\
	 & $i$ & Inclination of galactic plane \\
	\hline
	\multirow{3}{*}{Bulge} & $I_\mathrm{eff}$ & Intensity at half-light radius \\
	 & $r_\mathrm{eff}$ & Half-light radius \\
	 & $n$ & S{\'e}rsic index \\
	\hline
	\multirow{2}{*}{Disk} & $I_0$ & Central intensity \\
	 & $h$ & Exponential scale \\
	\hline
	Spiral arm & $I_0$ & Maximum intensity in the arm \\
    \noalign{\smallskip}
    \cline{2-3}
	\multirow{5}{*}{$r(\varphi)$} & $m_{0 \ldots 3}$ & Polynomial parameters for $\varphi \leq \varphi_\text{break}$ \\
	 & $l_{0 \ldots 3}$ & Polynomial parameters for $\varphi > \varphi_\text{break}$ \\
	 & $r_0, \varphi_0$ & Coordinates of beginning of the arm \\
	 & $\varphi_\text{break}$ & Azimuthal angle of arm break \\
	 & cw & Arm winding direction \\
    \cline{2-3}
	\multirow{4}{*}{$I_\parallel$} & $\varphi_\text{max}$ & Azimuthal angle of maximum intensity \\
	 & $\varphi_\text{cutoff}$ & Azimuthal angle of cutoff beginning \\
	 & $\varphi_\text{end}$ & Azimuthal angle of end of the arm \\
	 & $h_s$ & Arm exponential scale \\
    \cline{2-3}
	\multirow{3}{*}{$I_\bot$} & $w_e^\text{in}, w_e^\text{out}$ & Half-light distance inwards and outwards \\
	 & $n^\text{in}, n^\text{out}$ & S{\'e}rsic index inwards and outwards \\
	 & $\xi$ & Arm width increase rate \\ 
\noalign{\smallskip}\hline
\end{tabular}
\end{table}

In this section, we describe decomposition components and their parameters.
\par
We use the standard S{\'e}rsic function for the bulge component model \citep{1968adga.book.....S}:

\begin{equation}
\label{eq:Sersic}
I(r) = I_\mathrm{eff} \exp \left\lbrace-b_n \left[\left(\frac{r}{r_\mathrm{eff}}\right)^{\frac{1}{n}} -1\right]\right\rbrace
\end{equation}

It is parameterized by three quantities, namely the S{\'e}rsic index $n$, which describes the radial concentration of the bulge, the effective radius $r_\mathrm{eff}$, which encloses half of the light, and intensity at this radius $I_\mathrm{eff}$. The $b_n$ value is a normalization coefficient. The disk component was described by a simple exponential distribution

\begin{equation}
I(r)=I_0\exp\left(-r/h\right)
\end{equation}

with two parameters, its central intensity $I_0$ and exponential scale length $h$.
\par
We model each spiral arm independently with a function that has 26 free parameters. Almost the same model has been used in our first study devoted to the spiral structure of 29 galaxies from S$^4$G survey~\citep{2024MNRAS.527.9605C}, where it was shown that our model reproduces well the observed properties of spirals, their shape and light distribution. Here, we again describe the basic equations and parameters of our model and refer the reader to~\citet{2024MNRAS.527.9605C} for a detailed discussion of the reasons for the particular choice of functional form.

\par
The surface brightness distribution of an arm in polar coordinates $\left( r, \varphi\right)$ has a following form:

\begin{equation}
I(r(\varphi),\varphi) = I_0\times I_{\parallel}(r,\varphi) \times I_{\bot}(r - r(\varphi), \varphi)
\label{eq:model}
\end{equation}
We now provide details about each function from Eq.~\ref{eq:model}.
\par
Function $r(\varphi)$ describes the shape of the ridge-line of spiral in polar coordinates $(r, \varphi)$, where $r$ is galactocentric distance and $\varphi$ is the azimuthal angle. To define $r(\varphi)$, we use two polynomial functions of order $N = 4$ with polynomial coefficients $m_i$ and $l_i$. This is the only change of the model from~\citet{2024MNRAS.527.9605C}, where $r(\varphi)$ is described by a single polynomial. This change was done because the arms of M\,51 demonstrate clearly visible sharp bends inside them, which cannot be modelled properly with smoothly varying pitch angle. The arm starts at a radius $r_0$ and position angle $\varphi_0$, and hereafter $\varphi$ is counted from $\varphi_0$ along the direction of spiral winding, clockwise or counterclockwise, which depends on one more parameter and is determined visually. Spiral arms bend at $(r_\text{break}, \varphi_\text{break})$, where $r_\text{break}$ is determined from other parameters, including $\varphi_\text{break}$, to make both parts of the arm match. So $r_\text{break}$ is not an independent parameter in our model. Therefore, $r(\varphi)$ has a total of 12 parameters and is formally described as follows:

\begin{equation}
\begin{cases}
r(\varphi) = r_0 \times \exp \left(\varphi \sum_{i=0}^3 m_i\left(\frac{\varphi}{2 \pi}\right)^i\right),~~\varphi < \varphi_\text{break}\\
r(\varphi) = r_\text{break} \times \exp \left(\left(\varphi - \varphi_\text{break}\right) \sum_{i=0}^3 l_i\left(\frac{\varphi - \varphi_\text{break}}{2 \pi}\right)^i\right),~~\varphi \geq \varphi_\text{break}.
\end{cases}
\end{equation}
\par

The average pitch angle $\langle\upmu\rangle$ for any part of the arm can be easily found as the arctangent of the slope coefficient in a linear fit of the points of the spiral structure in the log-polar coordinates, i.e. $\log r = \tan \langle\upmu\rangle \times \varphi + \log r_0$. Pitch angle $\upmu$ at azimuthal angle $\varphi$ can be expressed as following:

\begin{equation}
\begin{cases}
\upmu = \arctan \left(\sum_{i=0}^3 \left(i + 1\right) m_i \left(\frac{\varphi}{2 \pi}\right)^i\right),~~\varphi < \varphi_\text{break}\\
\upmu = \arctan \left(\sum_{i=0}^3 \left(i + 1\right) l_i \left(\frac{\varphi - \varphi_\text{break}}{2 \pi}\right)^i\right),~~\varphi > \varphi_\text{break}
\label{eq:pitch}
\end{cases}
\end{equation}

\par
For $I_\parallel$, the region of growth ends at angle $\varphi_\text{max}$, reaching a maximum intensity. Exponential decline continues until $\varphi_\text{cutoff}$, and the cutoff region ends until $\varphi_\text{end}$. At $\varphi_\text{end}$, the flux reaches zero and the spiral arm ends. In exponential decrease regions, intensity falls with radius exponentially with scale $h_s$, and in cutoff regions, exponential decrease is also multiplied by a linear decrease between 0 and 1. Before $\varphi_\text{cutoff}$, $I_\parallel$ defined by this formula of 4 parameters:

\begin{equation}
I_{\parallel}(r(\varphi), \varphi) = \frac{1}{\bar{I}} \left(h_s \times \varrho(\varphi) \right)^{\varrho(\varphi_\text{max})} \exp \left( -\varrho(\varphi) \right)
\end{equation}

where $\varrho(\varphi) = (r(\varphi)-r_0)/h_s$ and $\bar{I}$ is a normalization constant to make the maximum of $I_\parallel$ equal to 1. The exact value of $\bar{I}$ is derived from the other parameters as $\bar{I} = (\varrho(\varphi_\text{max}) \times h_s) ^ {\varrho(\varphi_\text{max})} \exp(-\varrho(\varphi_\text{max}))$. As said above, at $\varphi_\text{cutoff} \leq \varphi \leq \varphi_\text{end}$, the function is multiplied by factor of $\left(1 - \frac{\varphi - \varphi_\text{cutoff}}{\varphi_\text{end} - \varphi_\text{cutoff}}\right)$.
\par
For the profile across the arm we implement a modified S{\'e}rsic function. This profile is fitted independently for direction $\rho$ toward ($\text{in}$) the galactic center and from it ($\text{out}$) relative to the arm peak, each with its own $w_e^\text{in/out}$ and $n^\text{in/out}$ parameters (and $b_n^\text{in/out}$, which depend on $n^\text{in/out}$). The parameter $\xi$ describes the widening of the arm towards the outer edge. The condition $\xi>0$ means that the spiral arm's width increases when traveling towards the edge of the galaxy, and  $\xi$ is the same for both in and out directions. 
In total, the profile $I_{\bot}(\rho,\varphi)$ is described by 5 parameters and has the following form:

\begin{equation}
\label{eq:I_bot}
I_{\bot}^\text{in/out}(\rho,\varphi) = \exp \left(-b_n^\text{in/out} \times \left(\frac{\rho}{\sqrt{w_e^2 + \left(\varphi \times \xi \right)^3}} \right)^{\frac{1}{n^\text{in/out}}}\right)
\end{equation}

\par
A centre point $X_0, Y_0$, inclination $i$, and positional angle PA are also amidst the parameters of each component. Centre point and inclination are usually the same for all components. When including these, the model for each spiral arm in M\,51 has 26 parameters. All parameters, including those for the galaxy plane's orientation, bulge, and disk are listed in Tab.~\ref{tab:parameters}. The rather large number of parameters is justified, since spirals are complex structures that are tricky to fit with a simple form. The advantage of the proposed model is that it is arranged relatively simply into individual subcomponents, and most of the parameters can be reliably guessed from the image. For example, $h_s$ can be guessed from the position of the spiral's end, which is equal to $3-5\, h_s$. Another benefit is that each arm is fitted separately, thus allowing for a fit even in non-symmetrical cases, which simpler models can not do. The model proposed in this Section can fit very difficult cases, including M\,51 as demonstrated below, and 29 galaxies from \cite{2015ApJS..219....4S} work~\citep{2024MNRAS.527.9605C}.

\section{Decomposition}
\label{sec:decomposition}

\begin{figure*}
\centering
\includegraphics[width=1.85\columnwidth, angle=0]{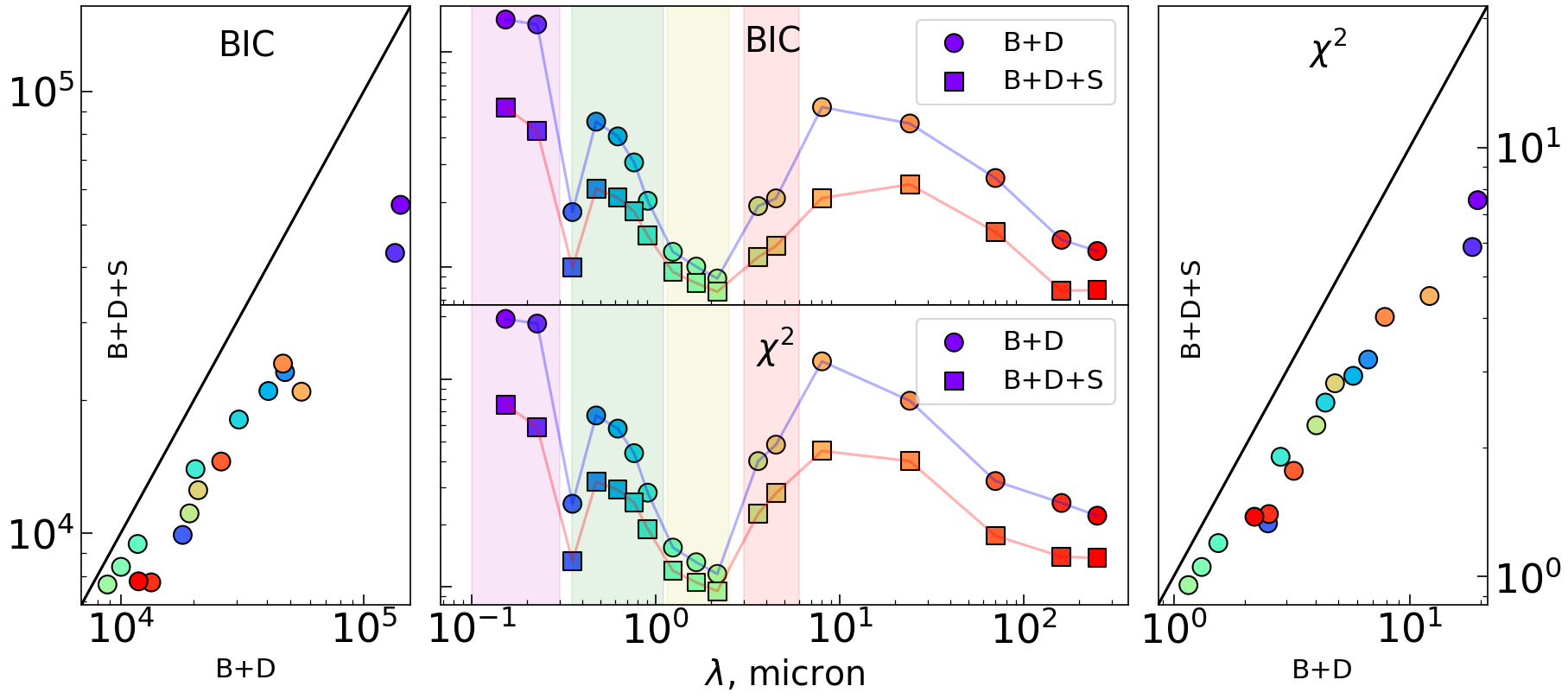}
\caption{$\chi^2$ (lower and right subplots) and BIC (upper and left subplots) comparison for images, convolved to the resolution of the 250~$\upmu$m image. The outside subplots show the comparison for the classical B+D versus B+D+S models, while the central subplots show the dependence on $\lambda$. The color of each marker represents its wavelength and it is the same among all Figures in this paper, where shown.}
\label{fig:bic}
\end{figure*}

\begin{figure*} 
   \centering
   \includegraphics[width=1.85\columnwidth, angle=0]{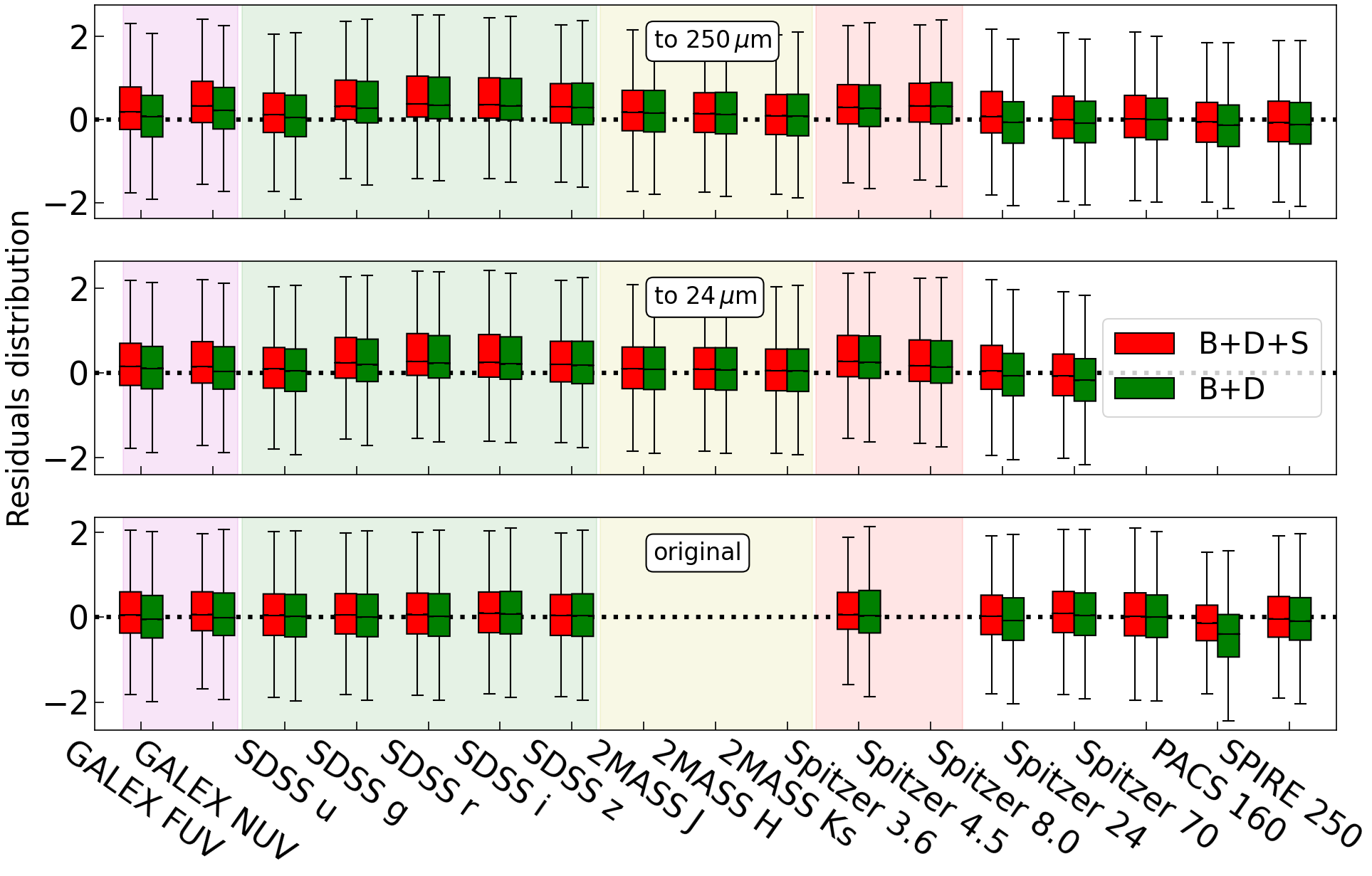}
   \caption{Distributions of the residuals for different models and bands, used in all three sets of images in this work. Each distribution is represented as a so-called boxplot, with the median, 25th, and 75th percentiles marked. For representation reasons, all residuals are normalizedin a way that interquartile range of B+D+S models (red-colored box, left model in each pair) equals to 1.}
   \label{fig:boxplots}
\end{figure*}

The model introduced in Sect.~\ref{sect:model} was implemented as new function in the {\small IMFIT} package \citep{2015ApJ...799..226E}. We ran the fitting twice with the same parameters for each image. For the first run, we applied the bulge plus disk model (B+D), while for the second run we used the same model, but with two additional spiral arms (B+D+S). 
\par
An important sidenote is that M\,51 contains an active galactic nucleus (AGN) at the center \citep{2016A&A...593A.118Q}. Its presence is manifested in the optical and other bands as the narrow central peak in Fig.~\ref{fig:slices} for images with the original resolution. Its appearance is mostly important in these images and it has a negligible effect on all structural parameters except those of the bulge. We include AGN in decomposition as a separate component only for the images in original resolution for these bands where it is visible, namely all SDSS bands and {\it Spitzer} 3.6, 8.0 and 24~$\upmu$m bands. We use a point source to model the AGN and find its contribution to the total luminosity of roughly 0.2\% in the optical, and the highest value of 0.5\% is achieved in the 24~$\upmu$m band.
\par
The main difficulty of approximating data with a complex model with many degrees of freedom lies in finding the proper initial values for the free parameters. To simplify fitting and speed up the optimization of the polynomial  $r(\varphi)$, we first traced the spiral arms' shapes manually, and then approximated the shape based on such measurements. Parameters describing the shape of arms were later used as initial guesses for fitting. We apply reasonable constraints to other parameters of spiral arms. For each image convolved to the 250~$\upmu$m-band image resolution and pixel scale, the fitting takes approximately half an hour for each {\small B+D+S} type of decomposition, while {\small B+D} model fitting took less than a minute for one band, with all available cores of Intel Core i5-7200U CPU.
We estimated the goodness of the fit through a visual comparison between real images and their models by analyzing residual images in Sect.~\ref{sec:validation}, and also via the bootstrapping procedure, described below.
\par
Some of the images in DustPedia have related error maps that are specifically prepared to account for various factors like image co-adding, dark noise, Poisson noise, etc. Unfortunately, such maps are not available for all bands. For example, the error maps are missing for the GALEX and SDSS bands. For these bands, we received error maps constructed by \citet{radiativeM51} (private communication) where the authors used the same approach originally presented in \citet{radiativeM81}. These were created from estimating the pixel-to-pixel noise, the calibration uncertainty, and the Poisson noise in the original data. For other bands, we use error maps provided in the DustPedia archive; for the {\it Spitzer} images we use original maps from the IRSA database\footnote{https://irsa.ipac.caltech.edu/}. 
These error maps are utilized in the $\chi^2$ calculation and to estimate parametric uncertainties in a correct and uniform way for all models.
\par
The orientation parameters, i.e. galaxy inclination $i$ and position angle PA, were fixed during decomposition. However, the choice of inclination for our modelling is a nontrivial task because the estimates of $i$ in the literature span a large range from $20\degr$ \citep{2020MNRAS.496.1610A,1999ApJ...523..136S} up to more than $50\degr$ \citep{2006MNRAS.367..469D}, and in some studies the inclination angle could not even be determined \citep{2009ApJ...697.1870E}. Such large inclination uncertainties are usually attributed to galaxies oriented close to face-on, but in the case of M\,51, its inclination has a large uncertainty due to the perturbed shape of the galaxy. It is also interesting that we find a similar range of $i$ uncertainty among individual bands if the orientation parameters are not fixed and we try to estimate them from fitting. To adopt a reasanoble inclination value for all bands under study, we decided not to choose it blindly from the literature and instead did the following procedure. We normalized the images in all 17 bands to the same intensity range, summed them up and performed decomposition of the resulting ``image'' treating $i$ and PA as free parameters. Although clearly unphysical, such a procedure produces the orientation parameters that work reasonably well for most of the images used: $i=46.2\degr$ and PA$=24.4\degr$, and that are supported by both kinematic \citep{2006MNRAS.367..469D} and photometric studies \citep{2010PASP..122.1397S}. We also tried to fit B+D models to all images with spiral arms masked off in the image, and the new measured orientation parameters were always close to the previously estimated values ($i=48.1\degr$ and PA$=18.5\degr$). We adopt the former pair of estimates for the orientation parameters.
\par
To estimate the uncertainties of the derived parameters and their stability, we apply a bootstrapping resampling procedure in {\small IMFIT}, running the decomposition many times. For each iteration, a new subsample of pixels is generated by sampling with the replacement from the original image and then fitted with the selected model. We ran 100, 500 and 10000 iterations in the FUV and near infrared (NIR) 2MASS $H$ band to find out how the number of runs affects the estimated uncertainty. We found that after 500 iterations, the uncertainties under consideration change by less than 10\% in both test bands. We thus consider this number of iterations for both models with and without spiral arms. Note that this was done only for images, convolved to 250~$\upmu$m resolution. We compute a standard deviation of the results for individual runs as an estimation of uncertainty. All uncertainties presented in Figures and Tables throughout the paper were calculated as such. Note that errors produced by {\small IMFIT} are usually small, as can be seen from Table~\ref{tab:bulge_disk} and Table~\ref{tab:spirals}, and thus in some cases are almost invisible in the Figures.  We also analyze the covariation matrix for different numbers of iterations, and do not find any significant unexpected degeneracy between parameters.

\section{Validation}
\label{sec:validation}

We test the quality and validity of the obtained decomposition models in this Section. Observed spiral arms are hard to fully represent by a smooth model due to their spurs, feathers, and star-formation regions \citep{2006ApJ...650..818L}. Moreover, the number and prominence of such additional features change with $\lambda$. For example, bands with significant income of young stellar population, such as GALEX FUV, contain many local clusters of ongoing star formation, which are not modeled here and contribute to residuals.
For this reason, we cannot directly compare the values of reduced $\chi^2$ in various bands with each other. For example, the best photometric model of \citet{2010AJ....139.2097P} with non-axisymmetric Fourier-based modeling of spirals for M\,51 provides  $\chi^2 = 54.2$ for the SDSS $r$ band compared to $\chi^2 \approx 23.4$ provided by our model. This does not mean that our model is better, because the models use different images and different numbers of parameters. A reduction of residuals after taking spirals into account is more important. As seen in Fig.~\ref{fig:bic}, for all three datasets in every case $\chi^2$ became smaller when we add the spiral arm into our model. 
\par
In order to check whether this improvement justifies the inclusion of the additional 26 parameters for each spiral arm, we use the Bayesian Information Criterion (BIC, \citealt{1978AnSta...6..461S,2017pbi..book.....B}), which applies a penalty term for the number of free parameters. First, we use the general form of BIC, which is $$\mathrm{BIC}=\chi^2 + k\cdot\ln{N}$$ where $N$ is the number of data points and $k$ is the number of free parameters in the model. We find that BIC is distinctly lower for our model with spiral arms for all bands. However, in this criterion all data points are assumed to be independent, which is generally not true for astronomical images due to the PSF. The fact that BIC might depend on correlations between adjacent pixels was discussed in \cite{2011ApJS..196...11S} and \cite{2014MNRAS.440.1690H}: the authors suggest modifying the formula to $$\mathrm{BIC}=\frac{\chi^2}{A_{\mathrm{PSF}}} + k\cdot\ln{\frac{n}{A_{\mathrm{PSF}}}},$$ where $A_{\mathrm{PSF}}$ is the number of pixels in the PSF and $n$ is the total number of pixels. We show $\chi^2$ and the adjusted BIC in Fig.~\ref{fig:bic}. We can see that all B+D+S models clearly demonstrate lower values than those without a spiral arms component. This justifies the increased complexity of our decomposition model.
\par
After applying a reliable model to an observed image, the residual image should contain only minor sources (such as star forming regions) and noise, with the pixels symmetrically distributed around a nearly zero median. For example, the Milky Way model in IR bands from \citet{2021MNRAS.507.5246M} demonstrates such properties for the residual image. In Fig.~\ref{fig:boxplots} we show distributions of the pixel values in the residual images for the models with and without spiral arms. It is not easy to show these distributions on the same scale due to the drastically different emissions at different wavelengths, thus we decided to normalize the distribution so that the interquartile range of B+D+S models (the red-colored box in Fig.~\ref{fig:boxplots}) equals unity. It is obvious that most of the medians of the residuals are close to zero for all of the bands and both of the model versions, which is indicative of good approximations. As the various percentiles show, the distributions are nearly symmetrical too, as expected. The symmetry is greater and the dispersion of the residuals is lower when spiral arms are taken into account. This alignment with our expectations validates all presented models in a very direct way. 

An important question to wonder about is, ``How do we know that spiral arms are fully extracted?'' For example, the model also will become better than disk plus bulge in terms of $\chi^2$ when we use the same arm model, but with less bright spirals, e.g. with $I_0$ intensity halved. Obviously, spiral arms will not be properly and fully attributed in such a situation. There are two pieces of evidence that show that we are doing it properly. Firstly, since the distributions of residual pixels shown in Fig.~\ref{fig:boxplots} are symmetrical and have peak values close to 0, all large-scale features, including spiral arms, should be properly subtracted. Otherwise, we would have seen that the distribution appeared skewed. Secondly, we could see that all ``bumps'' on radial profile images in Fig.~\ref{fig:slices} in all bands, which  appear due to spiral presence, are properly and fully attributed. 
\par
Another validating and stabilizing test is that close photometric bands should generally demonstrate similar parameters obtained from the decomposition, i.e. galaxy in the $u$ filter demonstrates intermediate properties between UV and optics. The Figures in the next sections show that this is indeed the case, and we thus consider this test to be passed. 
\par
Of course, decomposition of auxiliary images with a higher resolution, either original or convolved to the resolution of the 24~$\upmu$m image, and the following comparison of the results throughout Sec.~\ref{sect:discussion} additionally allow us to validate the findings obtained in this work.
\par
Lastly, another important confirmation of our analysis comes from the agreement between our decomposition results and the literature, demonstrated for the bulge in Sect.~\ref{sec:bulge} and for the disk in Sect.~\ref{sec:disk}. The work of \citet{2017A&A...605A..18C} is of special interest here, because authors measure disk radial scales for M\,51 in the same DustPedia bands, being careful to avoid the influence of spiral arms. A comparison of their results with ours in Sect.~\ref{sec:disk} shows the reliability of our work.

\section{Results and discussion}
\label{sect:discussion}

\begin{table*}
\centering
\caption[]{Parameters of the bulge and disk in the B+D and B+D+S models for images reshaped for 250~$\upmu$m resolution. Results for other two datasets listed in the Appendix.}
\label{tab:bulge_disk}
\begin{tabular}{l|c|c|c|c|c|l|l|c|c}
\hline\noalign{\smallskip}
Name & \multicolumn{3}{c|}{Disk} & \multicolumn{6}{c|}{Bulge} \\
 & \multicolumn{2}{c|}{$h$, arcsec} & $I_0^\text{B+D} / I_0^\text{B+D+S}$ & \multicolumn{2}{c|}{$r_\mathrm{eff}$, arcsec} & \multicolumn{2}{c|}{$n$} & \multicolumn{2}{c|}{$B/T$}\\
 & B+D+S & B+D & & B+D+S & B+D & B+D+S & B+D & B+D+S & B+D \\
\hline
\hline\noalign{\smallskip}
FUV & 88.5$\pm$1.1 & 79.1$\pm$0.5 & 2.70 & 21.7$\pm$0.2 & 11.1$\pm$0.0 & 0.04$\pm$0.00 & 0.01$\pm$0.00 & 0.07 & 0.00 \\
NUV & 76.8$\pm$0.9 & 69.0$\pm$0.1 & 3.25 & 25.6$\pm$0.5 & 13.5$\pm$0.0 & 0.11$\pm$0.01 & 0.01$\pm$0.00 & 0.11 & 0.05 \\
$u$ & 72.0$\pm$1.0 & 75.7$\pm$0.5 & 1.54 & 20.6$\pm$0.5 & 13.4$\pm$1.9 & 0.32$\pm$0.03 & 0.05$\pm$0.03 & 0.08 & 0.04 \\
$g$ & 70.0$\pm$0.7 & 72.8$\pm$0.4 & 1.40 & 16.1$\pm$0.3 & 11.2$\pm$1.1 & 0.63$\pm$0.03 & 0.05$\pm$0.02 & 0.08 & 0.05 \\
$r$ & 68.7$\pm$0.3 & 71.7$\pm$0.3 & 1.32 & 15.1$\pm$0.1 & 12.4$\pm$0.4 & 0.69$\pm$0.03 & 0.01$\pm$0.00 & 0.09 & 0.04 \\
$i$ & 70.2$\pm$0.5 & 72.8$\pm$0.3 & 1.29 & 16.0$\pm$0.2 & 12.7$\pm$0.4 & 0.91$\pm$0.04 & 0.26$\pm$0.00 & 0.10 & 0.08 \\
$z$ & 67.3$\pm$0.6 & 71.0$\pm$0.6 & 1.27 & 16.5$\pm$0.2 & 12.9$\pm$0.8 & 1.08$\pm$0.05 & 0.29$\pm$0.05 & 0.11 & 0.08 \\
$J$ & 59.8$\pm$0.8 & 65.6$\pm$0.4 & 1.17 & 16.8$\pm$0.4 & 13.5$\pm$0.4 & 1.11$\pm$0.09 & 0.36$\pm$0.00 & 0.13 & 0.10 \\
$H$ & 63.3$\pm$0.4 & 67.8$\pm$0.4 & 1.27 & 18.8$\pm$0.1 & 14.0$\pm$0.2 & 1.41$\pm$0.0 & 0.43$\pm$0.03 & 0.15 & 0.11 \\
$K_s$ & 60.9$\pm$0.9 & 66.2$\pm$0.4 & 1.25 & 18.2$\pm$0.3 & 14.4$\pm$0.2 & 1.15$\pm$0.06 & 0.45$\pm$0.03 & 0.15 & 0.12 \\
3.6~$\upmu$m & 71.0$\pm$0.5 & 70.7$\pm$0.4 & 1.60 & 20.5$\pm$0.2 & 15.3$\pm$0.4 & 1.13$\pm$0.04 & 0.40$\pm$0.05 & 0.17 & 0.11 \\
4.5~$\upmu$m & 75.6$\pm$0.7 & 72.5$\pm$0.4 & 1.74 & 21.0$\pm$0.2 & 15.9$\pm$0.4 & 1.06$\pm$0.04 & 0.39$\pm$0.04 & 0.18 & 0.11 \\
8.0~$\upmu$m & 57.5$\pm$0.0 & 60.1$\pm$0.6 & 2.13 & 25.9$\pm$0.0 & 18.2$\pm$0.6 & 0.09$\pm$0.00 & 0.05$\pm$0.00 & 0.13 & 0.07 \\
24~$\upmu$m & 67.6$\pm$0.3 & 62.3$\pm$0.4 & 2.62 & 26.6$\pm$0.2 & 19.2$\pm$0.5 & 0.12$\pm$0.00 & 0.05$\pm$0.00 & 0.21 & 0.15 \\
70~$\upmu$m & 68.8$\pm$0.5 & 62.5$\pm$0.3 & 4.10 & 24.5$\pm$0.3 & 17.0$\pm$0.8 & 0.32$\pm$0.01 & 0.05$\pm$0.00 & 0.22 & 0.14 \\
160~$\upmu$m & 67.3$\pm$0.5 & 64.5$\pm$0.3 & 2.51 & 26.1$\pm$0.3 & 19.9$\pm$0.6 & 0.20$\pm$0.02 & 0.05$\pm$0.00 & 0.18 & 0.11 \\
250~$\upmu$m & 77.3$\pm$0.5 & 68.7$\pm$0.4 & 2.93 & 23.0$\pm$0.3 & 17.9$\pm$0.7 & 0.09$\pm$0.00 & 0.05$\pm$0.00 & 0.11 & 0.10 \\
\hline
Mean & 69.6 & 69.0 & 2.01 & 20.8 & 14.8 & 0.62 & 0.17 & 0.13 & 0.09 \\
\noalign{\smallskip}\hline
\end{tabular}
\end{table*}

\begin{figure*}
   \centering
\includegraphics[width=1.85\columnwidth, angle=0]{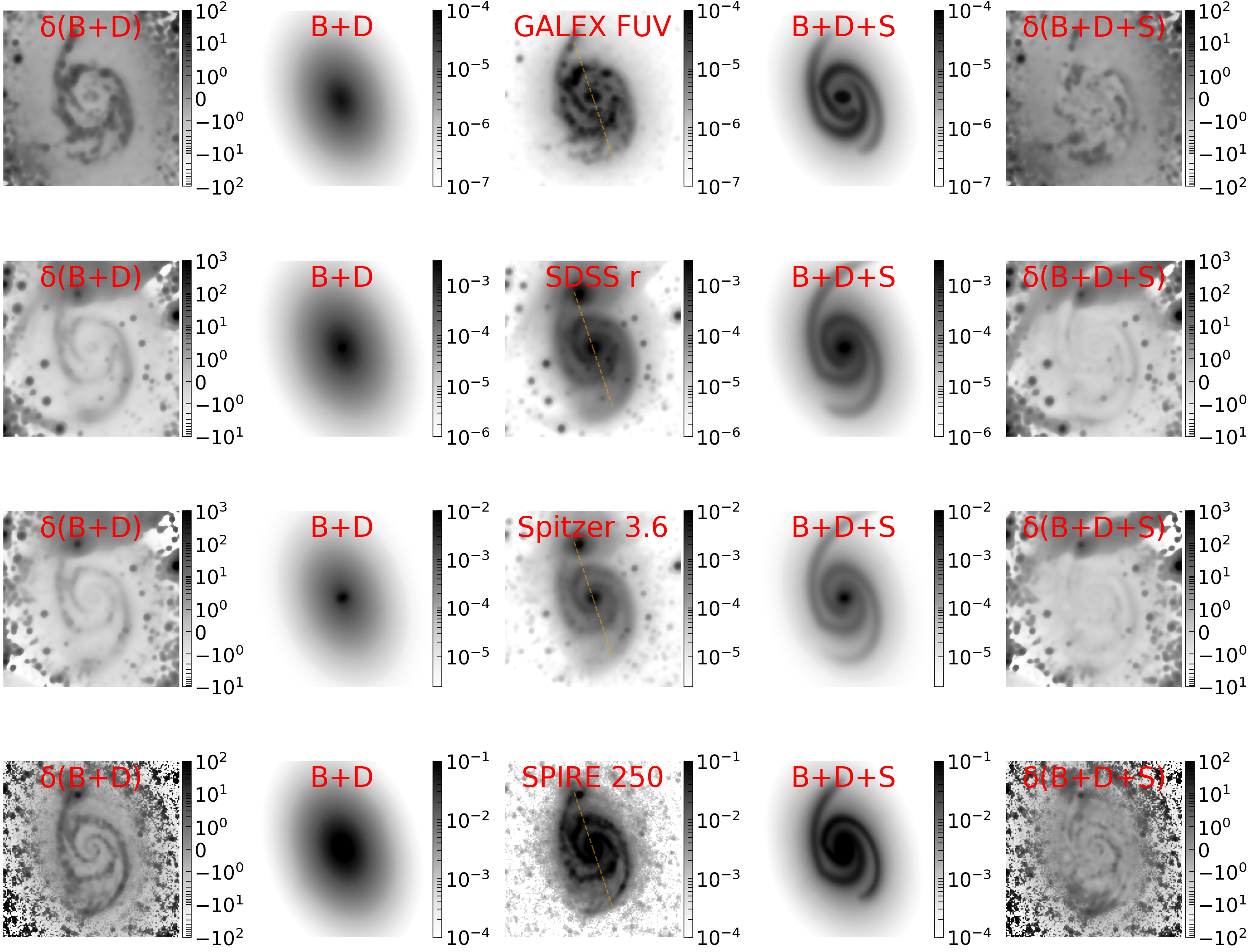}
   \caption{Original images (with the band marked), models and relative residuals for M\,51 in four different bands. The masks are not shown for a better comparison between the image and the model. The line on the central image marks the position of the PA.}
   \label{fig:main}
   \end{figure*}

\begin{figure*}
\centering
\includegraphics[width=1.85\columnwidth, angle=0]{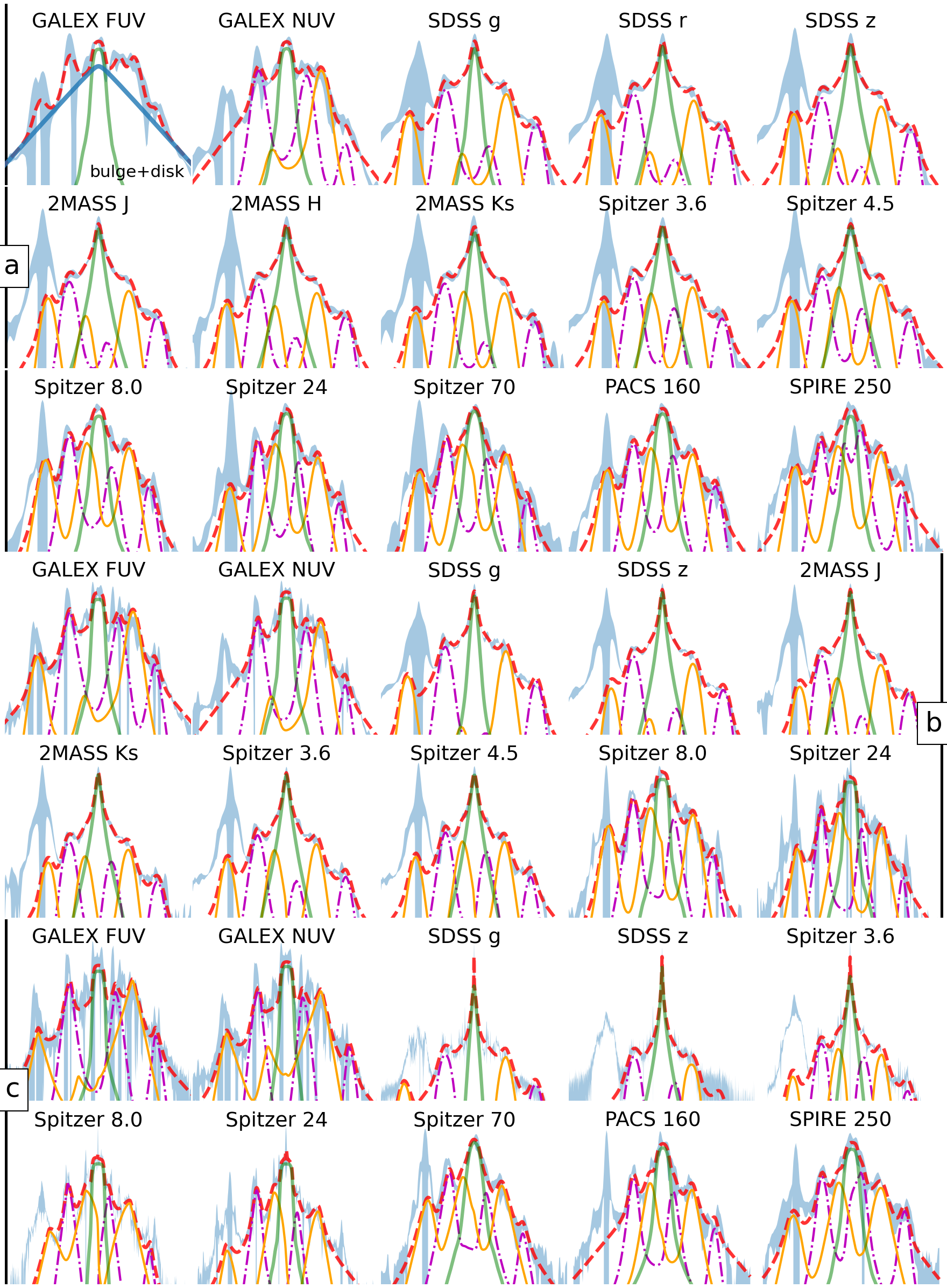}
\caption{Radial profiles along major axis for B+D+S models in three datasets (a - convolved to 250~$\upmu$m; b - convolved to 24~$\upmu$m; c - original resolution). The blue colour is the observed flux in log coordinates with its uncertainty (the left peak of the observed profile is due to NGC5195), the red dashed line shows the B+D+S model profile, green lines show bulge profile, other lines show profiles of spiral arms. Each arm represents by line with its own linestyle and color. The disk components are omitted for clarity, except for the GALEX FUV in the upper left corner, whose disk and bulge profiles are shown as an example. All profiles are normalized to have the same height of the central peak. Each horizontal axis spans in 7.5~arcmin in each direction, and omitted for visibility reasons. In data set (c) all SDSS and 3.6~$\upmu$m, 8.0~$\upmu$m and 24~$\upmu$m bands contains AGN as separate component, visible as an exceptionally narrow central peak.}
\label{fig:slices}
\end{figure*}

\begin{figure*} 
   \centering
   \includegraphics[width=1.95\columnwidth, angle=0]{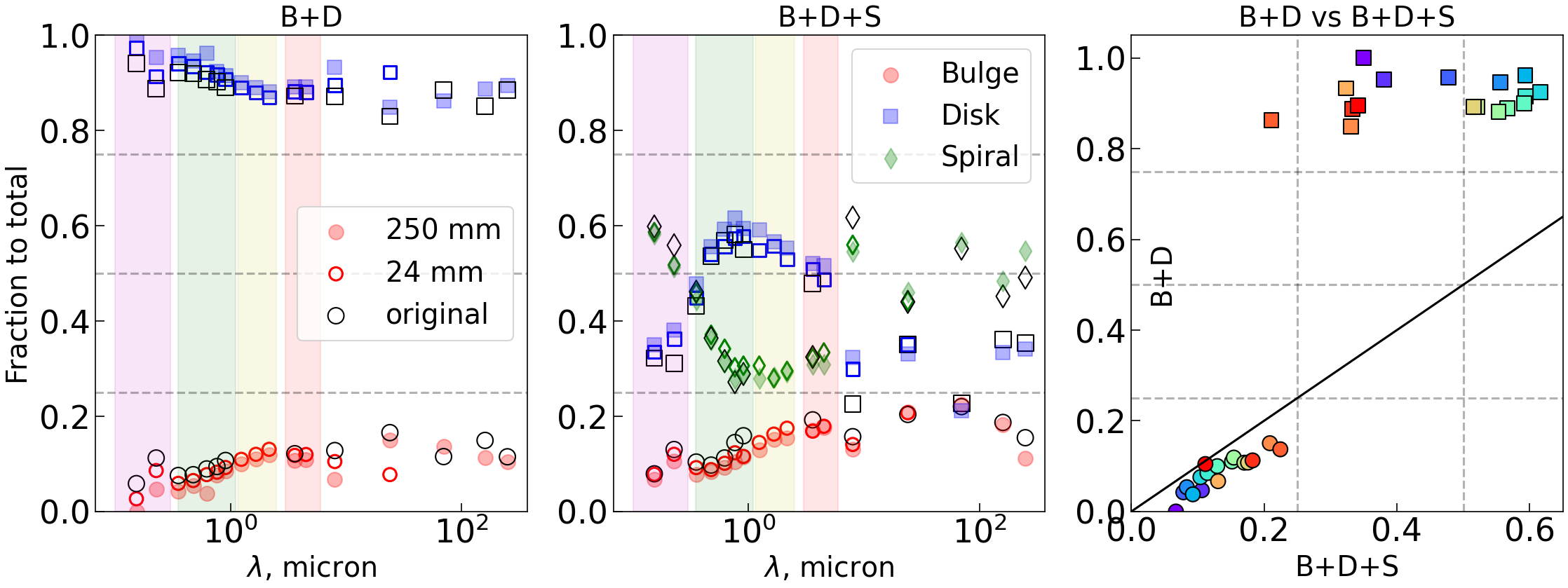}
   \caption{Light fractions in different components relative to total flux. The left plot shows the dependence of bulge-to-total ($B/T$) and disk-to-total ($D/T$) on wavelength $\lambda$ for B+D model. The central plot represents the same for B+D+S model. In these plots, filled symbols represent images convolved to 250~$\upmu$m, open colour symbols show images convolved to 24~$\upmu$m, black symbols mark images with the original resolution. The right plot compares bulge and disk fractions between models for main dataset, solid line marks 1:1 relation. } 
   \label{fig:frac}
\end{figure*}
   
In this Section we present the results of the fitting, a comparison with classical B+D models, and the effect that including spiral arms in M\,51 has on its components. Original images, fitted models, and the residuals for several bands are presented in Fig.~\ref{fig:main}. We do not show images for all 17 bands in the main sample, but the presented ones are selected from various parts of the spectrum spaced across the $\lambda$ range. Firstly, it is easy to see that the appearance of the M\,51 -- NGC\,5195 pair in the original images depends on wavelength, i.e. more flocculent with massive clumps in UV and FIR bands. Secondly, the disk and bulge sizes are slightly different. In the residual images for classical B+D models, the large unsubtracted spiral arms are easily distinguishable from the background. On the other side of Fig.~\ref{fig:main} are residual images do not demonstrate such large features. There is still plenty of emission left, especially in the UV and FIR images, but the remaining features are remarkably smaller and relate mostly to ``knots'' of star formation and arms' feathers. Finally, and most importantly, it is also easy to notice the good similarities between B+D+S models and the original image in all filters, and that properties of arms, including visible ``bends'', are fully and properly modeled.
\par
To get an intuition for how 1D profiles of spiral arms look in these models, we show slices along the major axis in Fig.~\ref{fig:slices}. We show slices for all three sets of models, but for visibility reasons some bands, which are similar to those displayed, are omitted. ``Bumps'' on these profiles, which arise from spiral arms, have different amplitudes in different parts of the spectrum. Models follow the profiles nicely to the full extent,  resulted not only in a good model of the spirals, but also for the disk and bulge regardless of the $\lambda$ considered. Since PA is fixed during the decomposition, each image in Fig.~\ref{fig:slices} shows a galaxy profile along the same photometric cut.
Note that the left part of the profile is significantly affected by NGC\,5195 presence even after masking, followed by a noticeable discrepancy between the model and data in this region.
\par
We present the light budget distribution across the bulge, disk and spirals in Fig.~\ref{fig:frac}. As this Figure states, different components occupy nearly the same level across considered wavelengths in the classical models. The disk component in B+D+S models share around 30---70\% with maximum in the optical and NIR bands. This maximum is expected because these bands are less affected by star formation in spiral arms. The contribution of the bulge component is maximal in bands constituted by the light of mostly old population, and contains 10---20\% of emission. This is similar to \cite{radiativeM51} results, where old stellar populations in the bulge represent roughly 15\% of the total emission. \citet{2010AJ....139.2097P} find a similar bulge fraction of 0.16 in their sophisticated model with a disk substituted of spiral arms, while \cite{2015ApJS..219....4S} find it as 0.13. We find that spiral arms are the most luminous part of M\,51 system, containing up to 60\% of light in UV bands and roughly half of the light in $\lambda > 4.5$~$\upmu$m parts of the spectrum. This is higher than a typical 20\%-40\% contribution of spiral arms of grand-design galaxies \citep{savchenko}. This is not an unexpected result since M\,51 is an outstanding example of a grand-design galaxy with spiral arms being the most distinct features.
The spiral arm contribution in M\,51 is maximal in the UV and IR due to the emission of young stars and re-emission by dust, respectively.
\citet{2018ApJ...862...13Y} use a large sample of 605 galaxies to measure spiral arms' strength in them using a discrete Fourier transformation, both in one and two dimensions in the $VRI$ optical bands, finding that the mean arm strength is systematically stronger toward bluer bands. \citet{2015MNRAS.446.4155K} perform a similar analysis and estimate that spirals' component amplitude is always lower in the $V$~band than at 3.6~$\upmu$m and 4.5~$\upmu$m.
\par
The contribution of spiral arms in the B+D+S versus B+D models is formed mainly due to the contribution of disk, while the contribution of bulge increases only slightly, as Fig.~\ref{fig:frac} states.
If spirals are not included in model, disk must be brighter to account for the light that is actually part of the spiral arms. This should also be true for central parts, where spiral arms are not present but the bulge is, and some light from the bulge is attributed to the disk in B+D model. This is the reason why the bulge-to-total ($B/T$) fraction increases in B+D+S model for all datasets. A similar conclusion about the bulge fraction was made in \cite{2014ApJ...780...69L} for NGC\,2748. Moreover, the extreme example as M\,51 clearly shows that if spiral arms are not included in decomposition, the obtained parameters, e.g. disk-to-total ($D/T$) ratio, will be incorrect in all parts of the spectrum since half of the light will be improperly attributed. Note, however, that this statement about $D/T$ overestimation relates to the axisymmetric (Fourier mode $m=0$) disk part only, because formal spirals emerge from the disk's material and should be considered as a piece of the disk, for example when estimating the $B/D$ ratio.

\subsection{Bulge parameters}
\label{sec:bulge}

A proper bulge decomposition for M\,51 is problematic because the size of the bulge is small\footnote{Probably, the feedback from the AGN should be the main reason for that \citep{2016A&A...593A.118Q}.}, and the bulge itself has a low S\'ersic index $n$. Furthermore, for our main set of images convolved to the 250~$\upmu$m resolution, the size of the bulge is comparable to the size of the PSF. This greatly affects the estimates of the bulge parameters, as we demonstrate below in the discussion of our results for the S{\'e}rsic index. \citet{2008MNRAS.384..420G} found that the structural properties of the bulge can be reliably retrieved if $r_{\mathrm{eff}}$ is larger than 0.8 times PSF HWHM (half of FWHM). Formally speaking, this criterion is fulfilled for the dataset considered here, and we should be able to estimate the bulge parameters without any issues. However, \citet{2008MNRAS.384..420G} obtained his criterion mainly for $n\gtrsim 1$, which is not valid for M\,51 (at least for some bands). 
\par
Moreover, the bulge area appears differently in different bands even in the original (not resampled) images. Specifically, in GALEX FUV and NUV bands, the bulge has a clumpy structure with sharp edges (see radial profiles in Fig.~\ref{fig:slices}, green lines), while the bulge radial profile is steep, but clearly shows gradual intensity decaying for longer wavelengths. However, same sharp edges also appear for Spitzer~8.0 and longer wavelengths. As we show below, bulges with sharp edges observed at short and long wavelengths have very low S\'ersic indexes $n$ (close to zero). We do not associate these unrealistically low indexes with either pixel resolution or PSF, but with the fact that we observe different structures or different parts of the same structure in such a wide range of wavelengths as we consider in the present work. We also note that the fact that two-component bulge+disc fits have complex degeneracies is well known \citep{2020ApJ...900..178L,2017MNRAS.466.1513R}, and it is indeed the case for M\,51, where ~\cite{2023MNRAS.526..118I} note that the bulge parameters are subject to significant uncertainty. In our study, we also note that the bulge parameters should be taken with caution, although some conclusions regarding the the bulge parameters can be considered reliable.

\par

\begin{figure*}
   \centering
  \includegraphics[width=1.95\columnwidth, angle=0]{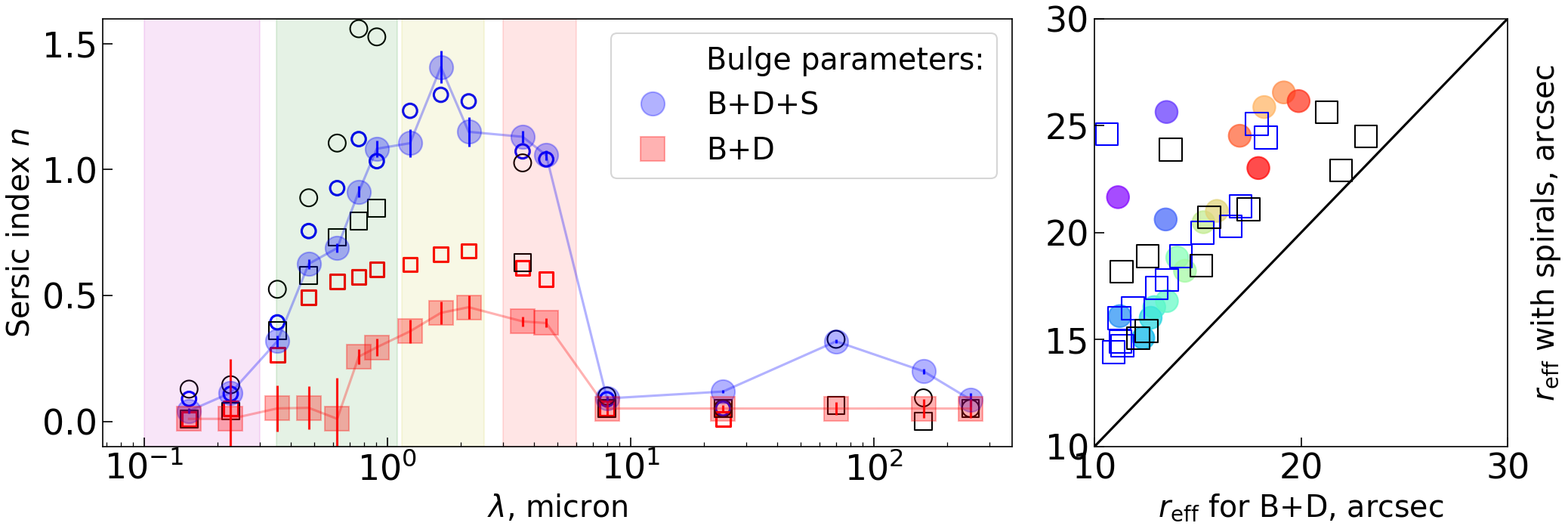}
   \caption{Bulge parameters obtained in decomposition. Left plot shows a dependence of S{\'e}rsic index $n$ on wavelength for both models. In the right plot, the effective radius $r_\mathrm{eff}$ is compared between two models.} 
   \label{fig:bulge}
\end{figure*}

The results of the bulge fitting are presented in Fig.~\ref{fig:bulge} and Table~\ref{tab:bulge_disk}. As can be seen, the effective radius $r_{\mathrm{eff}}$ is considerably larger for the B+D+S models in all cases. We link this to the fact that all the models without spiral arms suffer from an overestimation of disk brightness, as was discussed in Sec.~\ref{sect:discussion} concerning the difference of the $B/T$ fraction in the B+D and B+D+S models. Therefore, one can expect that the bulge will shrink to the very central area in a B+D model because the outer parts of the bulge will be fitted with a disk, and $r_{\mathrm{eff}}$ should decrease, accordingly. This is exactly what we see in Fig.~\ref{fig:bulge} (\textit{right} panel).
\par
In general, the value of $r_{\mathrm{eff}}$ lies within the 16--27~arcsec range, and is lower in the optical and NIR bands. Our estimates from Fig.~\ref{fig:bulge} agree with those from the literature: \cite{2015ApJS..219....4S} obtain $r_{\mathrm{eff}} = 21$~arcsec at 3.6~$\upmu$m, \cite{radiativeM51} provide 15~arcsec for the old bulge, and \cite{2008AJ....136..773F} measure $r_{\mathrm{eff}}=13.83\pm1.95$~arcsec in the $V$ band. Note that \cite{2008AJ....136..773F} also classify the bulge as a pseudobulge with nuclear spirals. The wavelength dependance of the bulge properties has been investigated only in the optical and NIR parts of the spectrum. In \citet{2014MNRAS.441.1340V}, \cite{2020A&A...633A.104P} and in other studies (see also table~3 in \citealt{2015MNRAS.454..806K}), authors find that $r_{\mathrm{eff}}$ decreases with wavelength, but only slightly. We do not find a similar trend. 
\par

The S{\'e}rsic parameter $n$ is small in M\,51 and could not be extracted reliably in the shortest and largest $\lambda$. In these bands, the model always has $n\approx 0$, as Fig.~\ref{fig:bulge} shows. Therefore, we conclude that $n$ is poorly determined from our data in some images. There are several reasons for that. First, the light distribution of the bulge in these bands actually looks very uniform and has steep edges. Such a profile is consistent with the S{\'e}rsic law with a very small $n < 0.25$. The exact value of the S{\'e}rsic  index is difficult to determine for such small values, since all bulges that have small $n$ look similar. In the 8.0~$\upmu$m band image in Fig.~\ref{fig:slices}, one can even see a dip in the centre of the surface brightness profile. Second, we observe $n \lesssim 1.5$ for all bands in the B+D+S model, and we find that $n$ decreases for the B+D model in every case. This is due to the same reason as for the $r_\mathrm{eff}$: the overestimation of the disk makes the edges of the bulge model steeper and lowers $n$ even more for the B+D models.
Finally, the fact that we convert original resolutions and PSFs to those in the SPIRE 250~$\upmu$m band has a huge impact on small structures such as the bulge in M\,51, as we see from the comparison of the different resolution images. In Fig.~\ref{fig:bulge}, it is clearly seen that models built for lower-resolution images have significantly lower S\'ersic index $n$ than for those with a higher resolution. The issue with the bulge fitting can be at least partially attributed to the seeing problems: as was shown in \cite{2001MNRAS.321..269T}, the parameters of the S\'ersic function can be severely affected by the PSF image smearing when the effective radius is comparable with the seeing HWHM. This problem is especially acute for PSF with extended wings~\citep{2001MNRAS.328..977T}, which we see in M\,51. In the decomposition pipeline used, we take the PSF into account, but its impact can still be high for a S\'ersic component whose effective radius is close to the image seeing~\citep{2008MNRAS.384..420G}.
This fact demonstrates that for some particular bands, the image parameters such as the PSF and resolution are crucial for accurate bulge fitting in M\,51.

\par

The $n$ values obtained are broadly consistent with the literature, considering a large scatter between the datasets with different resolution. Thus, \citet{1992A&A...257...85P} notice that the bulge in M\,51 is very small in the $B, V, R, I$, and $Z$ photometric filters. \cite{2015ApJS..219....4S} measure $n=0.995$ in the 3.6~$\upmu$m band (however they did not use any model to fit the AGN), \cite{radiativeM51} also find $n=0.995$ for the old bulge, and \cite{2008AJ....136..773F} finds $n=0.55\pm0.14$ for the $V$ band. As a rare glimpse into another part of the spectrum, \cite{2019A&A...622A.132M} apply a single S{\'e}rsic fitting for 320 galaxies from Dustpedia in {\it Herschel} bands between 100---500~$\upmu$m. They find that $n$ is almost constant within these wavelengths. It is typically below 1 and for some galaxies is even lower than 0.5, as their fig.~4 shows. This fact is also true for our analysis of M\,51 as Fig.~\ref{fig:bulge} states, but we note that it cannot be directly compared with \cite{2019A&A...622A.132M} because they do not consider a disk component. \citet{2010AJ....139.2097P} performs decomposition of M\,51 in $r$ band with and without spiral arms and finds that the S{\'e}rsic models for bulge have $n = 1.75$ for a model without spirals and $n = 0.67$ for a model with spirals. It's curious that in our results $n$ for a classical model resembles the value that~\citet{2010AJ....139.2097P} and is definitely lower than $n$ for the model with spirals. However, we note that they adopt two-S{\'e}rsic bulge for a model with spirals, with a secondary S{\'e}rsic model resembling AGN in its parameters, whereas they use only one-S{\'e}rsic bulge in a classical model, making $n$ for the bulge overestimated. Also, they use a S{\'e}rsic profile for the disk with $n$ much smaller than unity ($n = 0.33$ for classical model) and two-S{\'e}rsic profile with Fourier and bending modes for the disk in the model with spirals. This makes disk profile far from exponential and may cause the difference in estimated bulge parameters. Our considerations are supported by~\cite{2014ApJ...780...69L}, who use the same spiral models and find that the derived bulge parameters are very sensitive to both the specific details and the number of components used in the models. Considering the overall variation of the bulge $n$ with wavelength,~\citet{2023ApJS..267...26G} found a qualitatively similar trend for the M\,81 galaxy. At their figure~2, it's clearly seen that $n$ is smallest in the UV bands, increasing with wavelength and approaching a maximum at a few$\mu$m.

\subsection{Disk parameters}
\label{sec:disk}

\begin{figure*}
   \centering
   \includegraphics[width=1.95\columnwidth, angle=0]{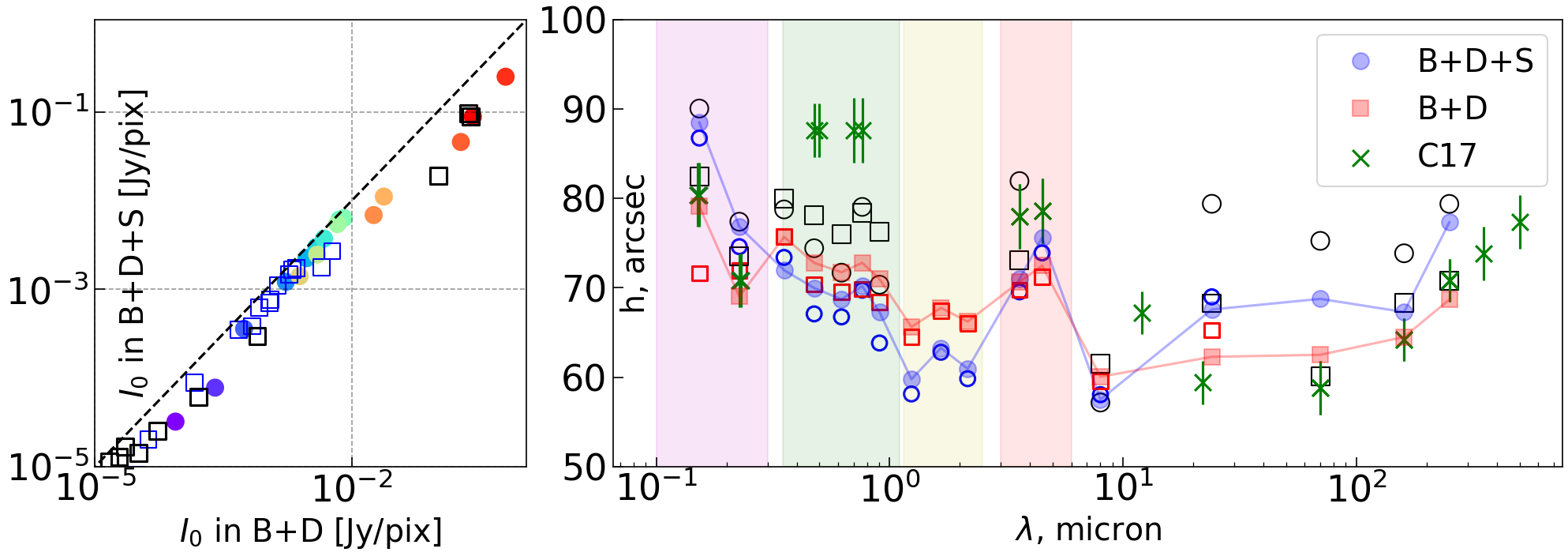}
   \caption{Disk parameters obtained in decomposition. In the left plot, the central intensity $I_0$ is compared between models without spiral arms (x-axis) and models with spiral arms (y-axis). The right plot shows the dependence of radial scale $h$ on wavelength for both models. Green crosses represent results obtained in \protect\cite{2017A&A...605A..18C}.} 
   \label{fig:disk}
\end{figure*}

\begin{figure}
   \centering
   \includegraphics[width=0.95\columnwidth, angle=0]{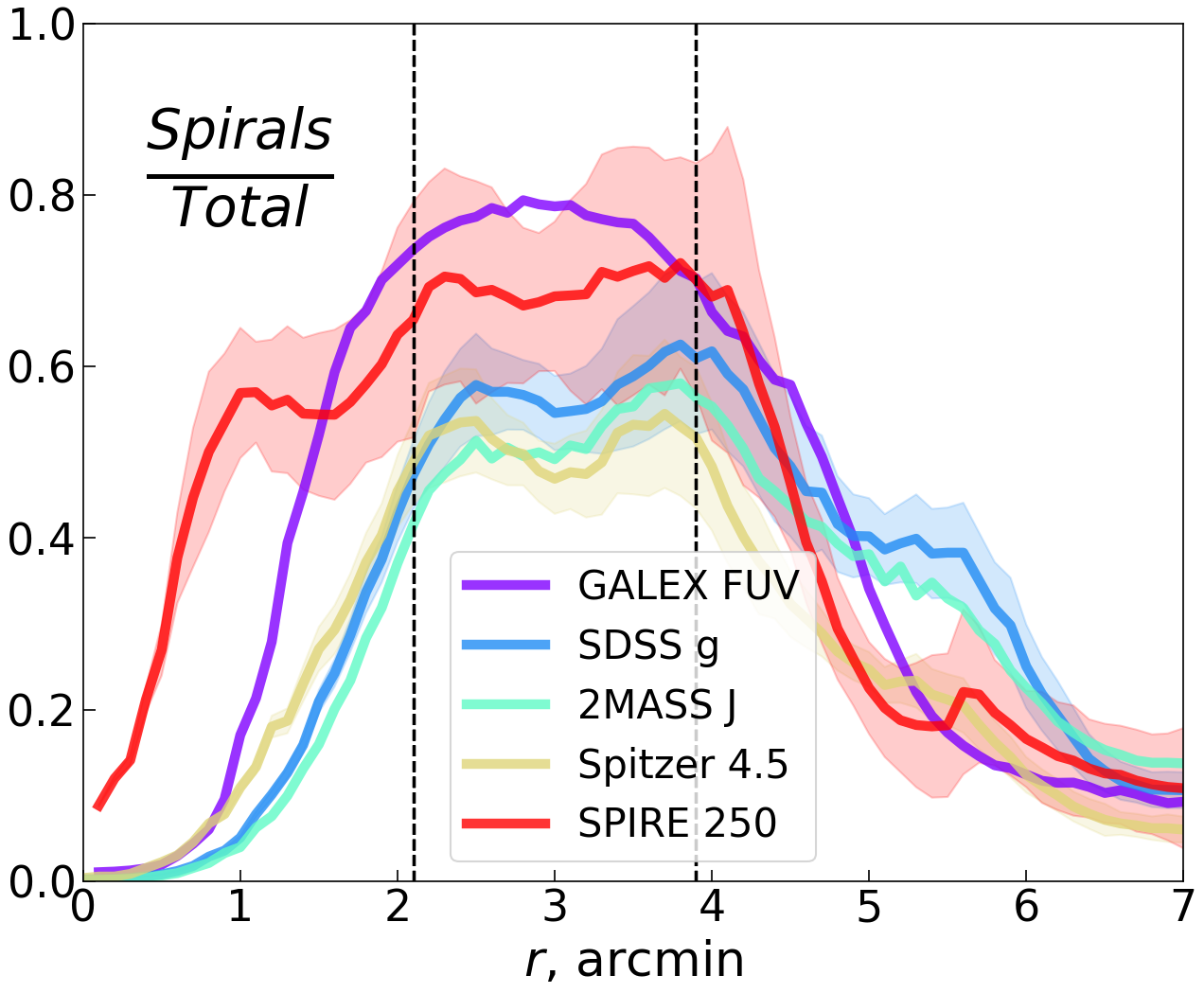}
   \caption{Fraction of modelled spirals to observed emission for azimuthally-averaged profiles. Vertical lines show limits used for disk scale fitting in \protect\cite{2017A&A...605A..18C}. Shaded regions represent uncertainties and for visibility reasons are only shown for a few bands.} 
   \label{fig:ST}
\end{figure}

As Fig.~\ref{fig:frac} demonstrates, the disk lost a large part of its luminosity after we included the spiral arms as a separate component, and thus a measurable change in its parameters is expected. The results regarding disk parameters are presented in Fig.~\ref{fig:disk} and in Table~\ref{tab:bulge_disk}. From the left panel of the Figure, we see that the disk's central intensity $I_0$ without spirals is larger than with spirals, as expected. In most cases, the changes are in a range from 25\% to a factor of three. It changes more at the ends of the analyzed spectral range, because the spiral arms' relative light budget there is higher, see Fig.~\ref{fig:frac}.
 \citet{2010AJ....139.2097P} finds the same effect of spiral arm inclusion in $r$ band, but with a smaller magnitude. 
\par

The shape of the curve for the obtained disk scale is presented in Fig~\ref{fig:disk}. This includes both the UV and MIR/FIR parts of the spectrum, and closely resembles the curve obtained in \citet{2017A&A...605A..18C,2019A&A...622A.132M,2020A&A...641A.119B}. In contrast to brightness, the disk exponential radial scale $h$ decreases from bluer optical bands to a minimum in NIR bands, which is 20\% smaller than maximum. Similar trends for $h$ were previously found in both the optical and NIR bands for early-type \citep{2010MNRAS.408.1313L} and late-type \citep{2012MNRAS.421.1007K} galaxies, and were confirmed by \citet{2014MNRAS.441.1340V}. We can see three $h$ peaks in both B+D and B+D+S models, which peaks are situated in the FUV, at 4.5~$\upmu$m, and at 250~$\upmu$m. This means that the youngest stars, oldest stars, and coldest dust respectively are distributed similarly throughout the interior of M\,51.
\par
Finally, the comparison shows that the exponential scales in the B+D+S models are systematically the in the FUV and FIR bands, and are similar in the optical and NIR. The results for other datasets demonstrate the same tendency. Even for such massive spirals as in M\,51, the effect is small, and the increase in $h$ when the spirals are taken into account in a modelling is 5--10\%. Curiously, the exponential scales for images with the original resolution are systematically higher, which can be related to the increase of the bulge S\'ersic index mentioned in the previous Section.

\par

In C17, the authors carefully investigated azimuthally-averaged radial profiles for these images, along with other DustPedia bands, in 18 face-on spiral galaxies including M\,51. This is of special interest, since they use the same images we use. 
Their measurements are shown as green crosses in Fig.~\ref{fig:disk}. We can see from the Figure that the radial scale $h$ is roughly the same between our data, except for optical filters. In C17, the authors do not fit the entire profile, but rather a small part of it, visible in their figure~A.1. Most of the 18 galaxies in C17 demonstrate exponential scales in optical bands that are lower than in UV (their figure~5), which is in agreement with what we find. 
We suspect that in optical bands, the slope of a smaller part of an azimuthally-averaged profile is probably not the same as the slope of exponential profile in larger parts of the disk. To test this, we conducted an analysis which is similar to C17 by masking the full image, except the area 2--4~arcmin which is considered in C17, and then fit only the exponential disk. We obtain results very similar to those in C17, and that directly confirms the hypothesis, and explains the observed discrepancy. 
Another way to explain the disagreement is seen in Fig.~\ref{fig:ST}, which shows the fraction of the modelled spirals to the observed emission in azimuthally-averaged profiles. We see that the range of radii used for fitting in C17 terminates in the middle of the second ``bump'', which is larger in optical and NIR bands, thus pushing the disk slope up and increasing $h$. The different azimuthally-averaged profile of spirals also explains the difference of the disk exponential scale between our B+D and B+D+S models, discussed above in this Section. This Figure also helps to understand another finding: 
in Fig.~\ref{fig:disk}, we see that reference values from C17 better agree with the B+D model without spiral arms. The reason is that C17 estimates $h$ from part of an azimuthally-averaged profile, which is located exactly where the contribution of the spiral structure into the signal is greatest, as Fig.~\ref{fig:ST} shows. Thus, the derived $h$ estimation is largely affected by the presence of spiral arms, resulting in a disk where arms' influence is not subtracted, i.e. B+D model. 

\par
   
Comparing with other works, \citet{2010AJ....139.2097P} find that the disk size in M\,51 with spiral arms as separate component becomes 25\% larger in $r$ band, which is larger than we measure, but consistent with the trend. Interestingly, in the \citet{2010AJ....139.2097P} model the disk's magnitude is different from ours by only 0.03~mag, which is strange taking into account that part of the disk light is in fact attributed to the spirals. The plausible reason for this is that they used a non-exponential model for the disk. \citet{1998AJ....116.1626B} do a classical decomposition in the $V$ band and estimate $h=107$~arcsec, which is larger than ours, but \cite{2008AJ....136..773F} find a much more consistent $h=86.8$~arcsec for that same filter. A precise decomposition at 3.6~$\upmu$m in \cite{2015ApJS..219....4S} results in an estimation of $h\approx84$~arcsec, which is close to ours, taking into account slightly different disk positional angle $39\degr$. Finally, it is important to note that spirals could also be responsible for necessitating a disk model change to broken form from a pure exponential \citep{2019A&A...625A..36W}, but this is not the case in this study.

\subsection{Spiral arms parameters}

\begin{table*}
\centering
\caption[]{Parameters of spiral arms for images convolved to 250~$\upmu$m resolution.}
\label{tab:spirals}
\begin{tabular}{l|c|c|c|c|c|c|c|c|c|c}
\hline\noalign{\smallskip}
       Band &  \multicolumn{2}{c|}{Spirals/Total} &  \multicolumn{2}{c|}{Pitch, deg} &  \multicolumn{2}{c|}{Width, arcsec} &  \multicolumn{2}{c|}{$h_s$, arcsec} &  \multicolumn{2}{c}{$h_s / h$} \\
$\,$ &  Arm1  &  Arm2  &  Arm1  &  Arm2  &  Arm1  &  Arm2  &  Arm1  &  Arm2  &  Arm1  &  Arm2  \\
\hline\noalign{\smallskip}
FUV & 0.29 & 0.30 & 9.1$\pm$0.2 & 11.5$\pm$0.5 & 40.5$\pm$2.4 & 44.0$\pm$2.8 & 66.7$\pm$2.3 & 15.7$\pm$1.4 & 0.75 & 0.18 \\
NUV & 0.26 & 0.26 & 9.4$\pm$0.2 & 12.1$\pm$0.4 & 38.9$\pm$6.4 & 44.6$\pm$2.2 & 77.1$\pm$3.7 & 17.2$\pm$1.3 & 1.00 & 0.22 \\
$u$ & 0.22 & 0.23 & 9.6$\pm$0.3 & 9.7$\pm$0.5 & 46.8$\pm$2.7 & 44.1$\pm$0.9 & 51.9$\pm$4.0 & 46.9$\pm$3.8 & 0.72 & 0.65 \\
$g$ & 0.19 & 0.17 & 11.0$\pm$0.3 & 14.1$\pm$0.5 & 47.0$\pm$1.6 & 45.8$\pm$0.7 & 51.0$\pm$2.6 & 62.4$\pm$2.8 & 0.73 & 0.89 \\
$r$ & 0.16 & 0.15 & 10.9$\pm$0.3 & 13.6$\pm$0.5 & 46.6$\pm$0.5 & 46.3$\pm$0.5 & 64.4$\pm$1.3 & 85.0$\pm$1.9 & 0.94 & 1.24 \\
$i$ & 0.15 & 0.13 & 11.1$\pm$0.3 & 13.7$\pm$0.3 & 46.9$\pm$0.7 & 46.9$\pm$1.3 & 66.0$\pm$1.5 & 102.5$\pm$2.8 & 0.94 & 1.46 \\
$z$ & 0.15 & 0.14 & 11.0$\pm$0.3 & 12.6$\pm$0.4 & 47.1$\pm$0.5 & 46.8$\pm$0.9 & 69.7$\pm$1.3 & 122.6$\pm$4.4 & 1.04 & 1.82 \\
$J$ & 0.16 & 0.12 & 11.5$\pm$0.3 & 15.5$\pm$0.3 & 48.0$\pm$0.6 & 46.9$\pm$0.9 & 57.0$\pm$1.9 & 59.9$\pm$4.1 & 0.95 & 1.00 \\
$H$ & 0.15 & 0.13 & 11.8$\pm$0.2 & 15.1$\pm$0.3 & 48.0$\pm$0.7 & 44.9$\pm$0.5 & 61.0$\pm$0.8 & 124.8$\pm$2.2 & 0.96 & 1.97 \\
$K_s$ & 0.15 & 0.14 & 11.7$\pm$0.3 & 16.3$\pm$0.4 & 48.0$\pm$0.7 & 43.7$\pm$0.5 & 57.5$\pm$1.3 & 156.5$\pm$3.1 & 0.94 & 2.57 \\
3.6~$\upmu$m & 0.16 & 0.14 & 12.0$\pm$0.3 & 15.7$\pm$0.5 & 42.5$\pm$5.8 & 41.0$\pm$0.6 & 53.6$\pm$1.4 & 122.2$\pm$2.5 & 0.76 & 1.72 \\
4.5~$\upmu$m & 0.16 & 0.15 & 12.0$\pm$0.3 & 16.0$\pm$0.5 & 40.7$\pm$4.8 & 37.7$\pm$0.5 & 49.4$\pm$1.8 & 129.3$\pm$2.3 & 0.65 & 1.71 \\
8.0~$\upmu$m & 0.27 & 0.28 & 12.4$\pm$0.3 & 14.7$\pm$0.4 & 36.2$\pm$0.0 & 41.1$\pm$0.0 & 36.4$\pm$0.0 & 173.2$\pm$0.0 & 0.63 & 3.01 \\
24~$\upmu$m & 0.23 & 0.23 & 13.4$\pm$0.3 & 15.4$\pm$0.4 & 24.3$\pm$0.8 & 33.9$\pm$0.9 & 26.8$\pm$1.1 & 72.1$\pm$1.5 & 0.40 & 1.07 \\
70~$\upmu$m & 0.28 & 0.29 & 14.5$\pm$0.2 & 14.8$\pm$0.3 & 27.6$\pm$2.2 & 38.4$\pm$3.3 & 20.4$\pm$1.9 & 131.2$\pm$3.1 & 0.30 & 1.91 \\
160~$\upmu$m & 0.25 & 0.23 & 16.5$\pm$0.3 & 14.5$\pm$0.4 & 36.8$\pm$1.8 & 38.6$\pm$1.8 & 32.5$\pm$4.4 & 144.8$\pm$4.5 & 0.48 & 2.15 \\
250~$\upmu$m & 0.31 & 0.24 & 14.8$\pm$0.3 & 14.0$\pm$0.4 & 18.8$\pm$0.7 & 38.8$\pm$1.8 & 114.6$\pm$2.4 & 176.1$\pm$5.4 & 1.48 & 2.28 \\

\hline
Mean & 0.21 & 0.20 & 11.9 & 14.1 & 40.3 & 42.6 & 56.2 & 102.5 & 0.80 & 1.52 \\
\hline
\end{tabular}
\end{table*}

Since the introduced spiral arm model consists of three parts, it is natural to also present the decomposition results using them: the shape, the profile along the arm, and the profile across the arm. All parameters discussed below for both arms are also listed in Table~\ref{tab:spirals} for the main dataset, and in Appendix for other models. It is important to mention that the arm pointed to the satellite galaxy NGC\,5195, which we called Arm 2 hereafter, is harder to constrain because its end, which is most likely a material feature \citep{2017ApJ...845...78C}, is located in a masked area and thus could vary a lot between images.
\par

\begin{figure*}
\centering
\includegraphics[width=1.75\columnwidth, angle=0]{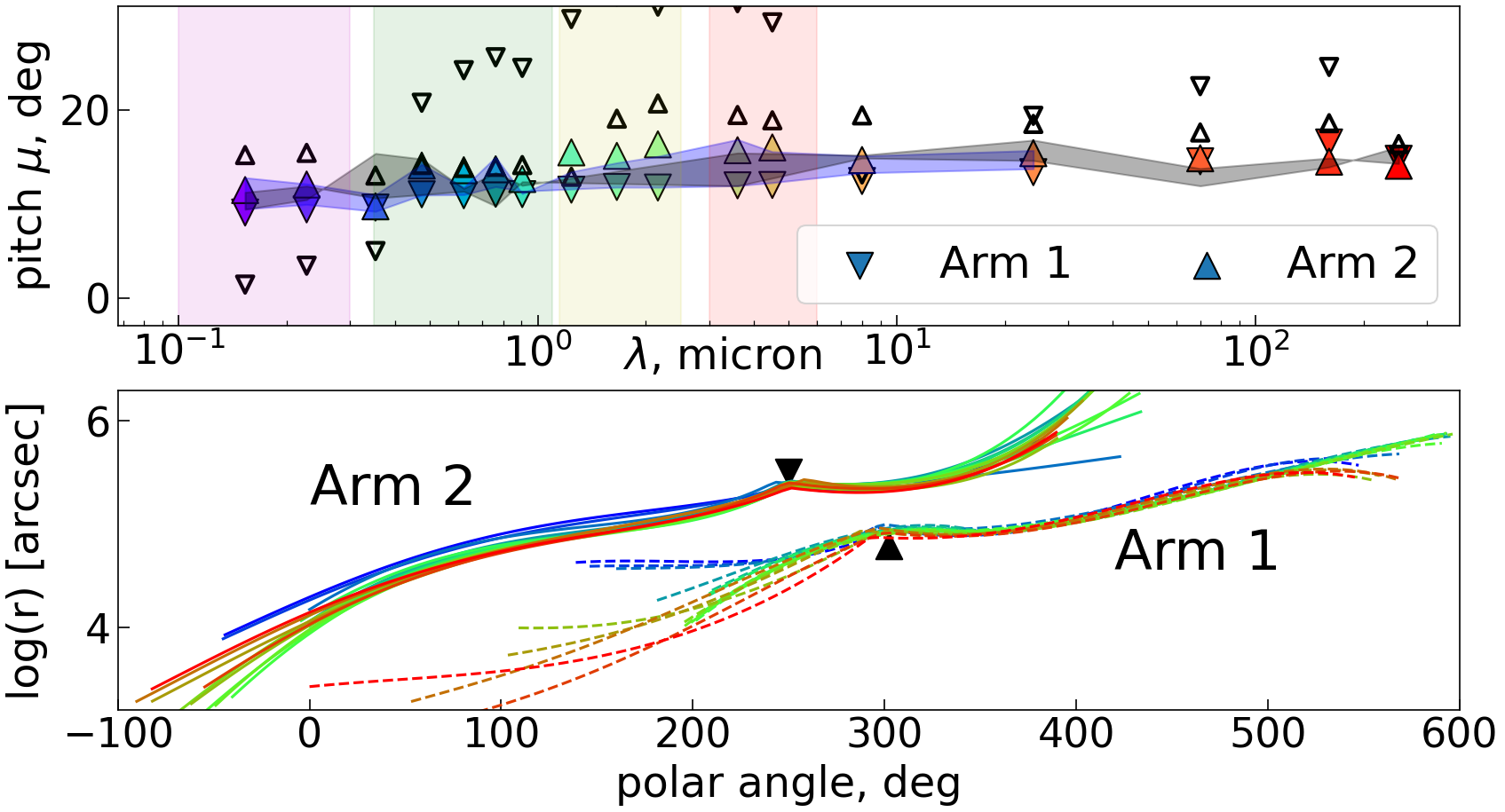}
\caption{The upper plot shows pitch angle measurements for both arms, and the bottom plot shows spiral arms in the log-polar coordinate system. Filled symbols are for the pitch angles measured in the main dataset using the arm's full extent, while smaller open symbols show pitch angle measured for the arm's part from the beginning till the ``bend'', marked in the bottom plot. Filled area mark pitch angle results for both arms in data convolved to $24~\upmu$m (blue) and with original resolution (grey).}
\label{fig:pitch}
\end{figure*}

\subsubsection{Pitch angles}

The pitch angles $\upmu$ are probably the easiest property of the spiral arm to measure, hence there broad usage in the literature related to this topic. However, there are several reasons for why its measurement is a harder task than it looks. Firstly, it is a relatively rare situation when $\upmu$ is measured for individual arms instead of averaging for the whole pattern, which the most common Fourier-based technique does \citep{2018ApJ...862...13Y}. Secondly, pitch angle is not constant along the arm \citep{1981AJ.....86.1847K,2013MNRAS.436.1074S}. This is easy to see for M\,51 in Fig.~\ref{fig:pitch} and directly in observational images, because both arms change their angle substantially. Hence, it is important to specify the range of measurement. Third, it is usually an ambiguous task to trace arm peaks to fit a logarithmic curve. 
To solve this issue, \citet{Tamburro2008} and other authors use cross-correlation to find a better motivated difference between arms in different $\lambda$, taking into account not only peaks, but the shape of the arm around the peak. Our approach to modelling the arm seems to be even more reasonable, because our pitch angle estimation takes into account the whole light distribution along the arm's body.
\par
Galaxy M\,51 has visibly asymmetrical arms with different shapes and pitch angles, shown in Fig.~\ref{fig:pitch}. Different $\upmu$ values are presented in the literature. \citet{2014AJ....148..133P} determined in M\,51 for Arm 2 $\upmu=19\degr$ using $8~\upmu$m images (see also \citealp{2023MNRAS.524...18M}). In \citet{2001ChJAA...1..395M}, the pitch angles were determined for each arm separately using DSS images, and the resulting values are $16.7\degr$ and $15.8\degr$. \citet{1982ApJ...253..101K} measured pitch angles in M\,51 using H{\sc ii} regions, estimating a range of $13-23\degr$. \citet{2015ApJ...800...53H} also used positions of 800 H{\sc ii} regions in M\,51 and measured that the pitch lies in the range of $8-13\degr$. \citet{2013ApJ...762L..27H} fitted logarithmic spirals in the arms of M\,51 using the SDSS $i$-band image and got a pitch angle slightly above $17\degr$. \citet{2007ApJ...665.1138S} used CO observations to determine a pitch angle of $21.1\degr$. \citet{2011MNRAS.414..538K,2015MNRAS.446.4155K} measured $13-14\degr$ for a range of NIR and optical images. For a subsample of galaxies visible in $3.6~\upmu$m, \citet{2019A&A...631A..94D} measured pitch angles using the Fourier method and obtained $17.1\degr$ for M\,51. Finally, \citet{2020MNRAS.496.1610A} determined pitch angles which lie within the range from $9.6\degr$ to $12.9\degr$ for M\,51 in a broad set of GALEX FUV, $B$ band, $3.6~\upmu$m and $8~\upmu$m bands. 
\par
The quite significant dispersion of the pitch angle values, taken from the literature, and the lack of agreement among them show the difficulty associated with accurate pitch angle measurements, as mentioned previously in this Section. Obviously, many previously measured values lie between the range $8\degr$ and $23\degr$ for individual arms because commonly accepted techniques like the 2DFFT average the whole pattern. In Fig.~\ref{fig:pitch}, we show pitch angles, measured using the whole arm extent. We obtain $\upmu\approx 10\degr-12\degr$ for both Arm 1 and Arm 2 that are consistent with those from other works. All data sets give similar values. Measured $\upmu$ values are increases with wavelength slightly. 
The discrepancy of the literature estimates may be due to the fact that different parts of arm is used to measure $\upmu$ in different studies. To show this, we also find $\upmu$ for parts of the arms before the bending ``knee'' visible in Fig.~\ref{fig:pitch}, where the pitch angle value changes abruptly. For its estimation, we use part of each arm that occupies roughly $200\degr$ of azimuthal angle. It is easy to see that the resulting values are considerably higher than the pitch angle measured for the full extent of the arms, reaching up to $30\degr$. This fact explains why previously measured $\upmu$ values could exceed $20\degr$ if a different part of the arm is used for measuring the pitch angle.

\par

\subsubsection{Density wave presence}

The problem whether or not M\,51 posses long-lived density wave is still debated despite the obvious influence of the tidal forces. If the galaxy exhibits a long-lived density wave for the whole spiral pattern, the pitch angles in bands associated with newborn stars, e.g. at UV, $8~\upmu$m and $24~\upmu$m wavelengths, should be lower than angles associated with the main bulk of old stars at $3.6~\upmu$m and 4.5~$\upmu$m (for example, see \citealt{2018ApJ...869...29Y}). The evidence for and against this picture in M\,51 is numerous and is roughly divided in half \citep{2020NewA...7601337V}. The fact that we do not observe such angles' inequalities in our models should count as an independent argument that M\,51 does not maintain a long-lived density wave for the whole spiral pattern \citep{2010MNRAS.403..625D,2018MNRAS.478.3590S,1982ApJ...253..101K}. It is important to note that, unlike previous studies, our finding is based on the total distribution of light across the arms. Since pitch angle measurements depend on where they are measured, as demonstrated in Fig.~\ref{fig:pitch}, we also estimate $\upmu$ angles over the entire extent using eq.~\ref{eq:pitch} and an additional window function approach previously used in \cite{2013MNRAS.436.1074S}. The conclusion remains the same, except probably for the tips of both arms, which are expected to be less constrained.
\par
The natural question, given the limited resolution used, is whether the effect is measurable in principle. We could roughly estimate its magnitude. There is no clear consensus in the literature on the location of the corotation radius in M\,51, which in some studies is found at 230~arcsec \citep{1982PhDT........35M} or at 300~arcsec \citep{2003AJ....126.2317O} from the center, or it may even have several corotation radii \citep{2008ApJ...688..224M}. However, in most works it is measured to be within 110--160~arcsec \citep{Tamburro2008,2013MNRAS.428..625S,1995ApJ...445..591E,2020MNRAS.496.1610A,1989ApJ...343..602E,2000MNRAS.319..377S}. We assume that the pitch angle is $12\degr$, corotation radius is 2.5~arcmin, length of the spiral is 5~arcmin, and rotation velocity is constant $v=215~\mathrm{km\,s^{-1}}$ \citep{2011AJ....141...24V}. Then we explore how the pitch angle of an arm in different band changes if we assume time travel being $\Delta t$ between those bands, e.g. between UV and $3.6~\upmu$m. For $\Delta t = 20/50/100$~Myr, the  variation in pitch angle is $\upmu > 1\degr/3\degr/4\degr$, accordingly. Therefore, for $\Delta t > 50$~Myr, the pitch angle change within the given assumptions is larger than the difference between $\upmu$ for the arms and its errors. Hence, associated pitch change should be in principle detectable e.g. between UV bands, where most of the stars are approximately 100~Myr old, and $3.6~\upmu$m band, where stars substitute most of the mass. However, the question of which offsets between wavebands in spiral arms are expected and in which direction is more puzzling than it looks \citep{2016ApJ...827L...2P,2019ApJ...874..177M,2023MNRAS.524...18M}. One should take this 
 into account, along with the signs of different nature of the outer parts of the arms \citep{1974ApJS...27..449T} and limited knowledge of `pixellation' and PSF influence (see, however, figures for M\,51 in \citealt{2023MNRAS.524...18M}). Hence we conclude only that M\,51 does not demonstrate signs of a long-lived density wave for the spiral pattern as a whole. Nevertheless, the implemented approach of the arms modelling may yield more definitive results about the nature of spirals in other galaxies.


\subsubsection{Profile along the arm}

The profile along the spiral arm is described in Sect.~\ref{sect:model} by several angles, including the position of the maximum, and by radial scale $h_s$. We notice that the beginning of the spiral $r_0$ shifts towards the center of the galaxy at larger $\lambda$, becoming 0.5~kpc closer in the FIR bands than in the UV. Note that this is evident for the FUV image in Fig.~\ref{fig:main} and Fig.~\ref{fig:pitch}, where the arms are visibly separated from the center. This may be explained by the fact that spiral arms are much more ``patchy'' in the UV, so $\chi^2$ becomes better if we omit several star formation areas located close to the center.
The angle $\varphi_\text{cutoff}$, defined as the location of change from the exponential to linear slope, is not constrained well enough. The same is true for the length of the linear segment. The reason for this is that the behavior of $I_{\parallel}(r)$ is close to a linear function for some parameters even before the cutoff point, thus allowing us to move the point in any direction without changing significantly the resulting profile.

\par
\begin{figure*}
\centering
\includegraphics[width=1.95\columnwidth, angle=0]{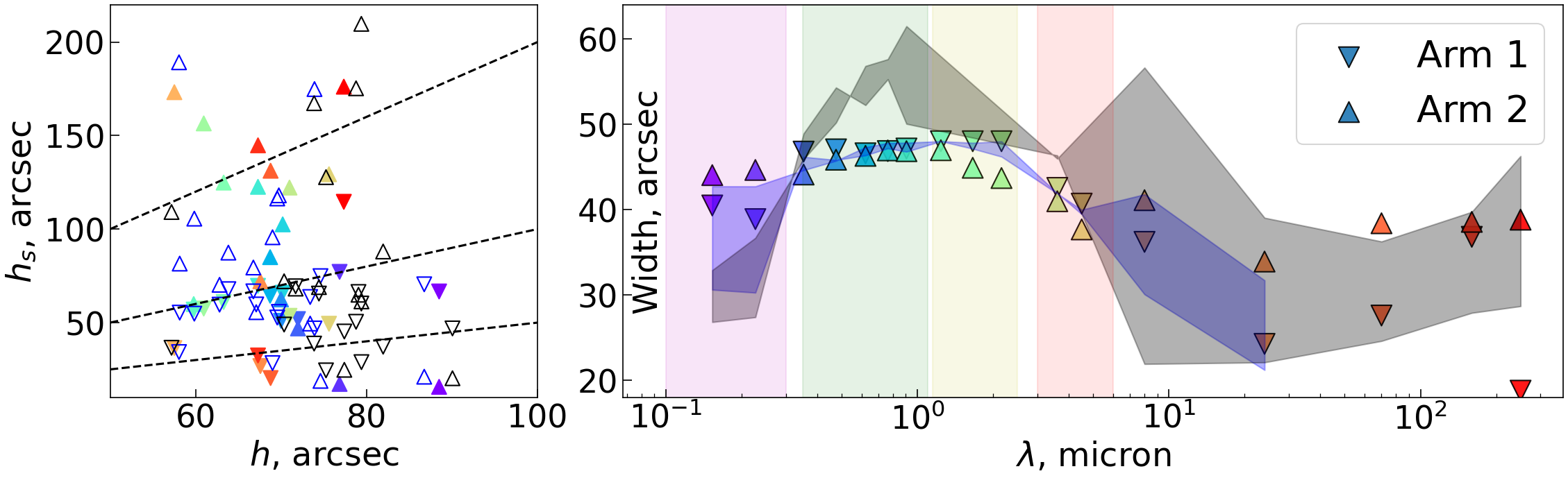}
\caption{Spirals' radial scales $h_s$ and widths, obtained in decomposition. The left plot shows the radial scale $h_s$ compared to the disk radial scale $h$, dashed lines show $0.5x$, $1.0x$ and $2.0x$ dependencies. Down-pointing triangle mark Arm 1, and triangle mark Arm 2, models are colour-coded as those in Fig.~\ref{fig:disk}. The right plot shows the dependence of spiral width on wavelength for both arms, coloured areas similar to those in Fig.~\ref{fig:pitch}.} 
\label{fig:width}
\end{figure*}

The radial scales of the arm $h_s$ roughly follow the same distribution as the arm width in Fig.~\ref{fig:width}, but for the Arm~2 it is significantly larger in most bands. This fact is partially noticeable in Fig.~\ref{fig:main}, where the flux of the arm connected with the companion NGC\,5195 remains relatively the same, but rapidly decreases towards the disk's edge for the Arm~1. The ratio of the arm and disk scales $h_s/h$ also has the same shape for both arms, and its value lies in a range of 0.5---1.5, as shown in the left part of Fig.~\ref{fig:width}. This is a typical range of values for this parameter, as our study of the spiral arms in 29 galaxies at $3.6~\upmu$m demonstrates~\citep{2024MNRAS.527.9605C}. In cases where the $h_s/h$ value is larger than unity, we should expect that the spiral arms' brightness relative amplitude increases towards the edge of the galaxy, which is observed in real galaxies, decomposed with the Fourier method \citep{2011MNRAS.414..538K,2015MNRAS.446.4155K}.
It is interesting to note that the values of $h_s/h$ drop in MIR bands, and this ratio rises then almost to the maximum level in the $250~\upmu$m band, which should somehow relate to the processes that govern the hot/cold dust distribution. All the mentioned properties of radial scales also closely resemble those in two auxiliary datasets of images.
\par
In Fig.~\ref{fig:ST}, we show which part of an azimuthally-averaged profile the spirals occupy. In all parts of the spectrum the picture is nearly the same: rise until 2~arcmin, then plateau with two noticeable peaks, until 4~arcmin, then decline with increased uncertainties. The spirals-to-total ratio in the plateau is equal to 0.5--0.8 and is greatest at the UV and IR ends of the spectral range used. The first and second peaks have nearly the same heights. These findings are supported by numerous works by other authors. For example, \citet{2011MNRAS.414..538K} fits in azimuthal profiles for  M\,51 in optical, $3.6~\upmu$m and $4.5~\upmu$m bands with a sine wave, obtaining spiral relative amplitudes. They clearly demonstrate a drop of luminosity between peaks, similar to our findings. \citet{1992ApJ...385L..37K} studied star formation efficiency in the spiral arms of M\,51, and also clearly demonstrates two peaks on contrasting images for individual arms, as does  \citet{1989ApJ...343..602E} for images in $B$ and $I$ bands. \citet{1989ApJ...343..602E} state that these modulations may be signs of interfered patterns. In \citet{2010PASP..122.1397S} their figure~13 clearly shows similar picture for $3.6~\upmu$m image. \citet{2009A&A...494...81S} and \citet{2017ApJ...845...78C} studied the spatial distribution and ages of star clusters in M\,51 relative to its spiral pattern. They identified two peaks in the profile of star formation and in the gas surface density in the radial direction, placed at 2.5~kpc and 5~kpc. The gap in the arms' luminosity is also clearly visible in the smoothed 2MASS H band image in figure~7 from \citet{2009A&A...494...81S}. \citet{2017ApJ...845...78C} noticed that all of the clusters' distributions are located in the centre of the spiral arms, and that its peak height decreases with cluster age, which is qualitatively in agreement with Fig.~\ref{fig:ST}. In \citet{2014AJ....148..133P}, the authors study $8~\upmu$m images of M\,51 and clearly show  two peaks of arm amplitude for an $m=2$ Fourier component in their figure~2. \citet{2013MNRAS.433.1837V} rigorously study spiral structure in M\,51 using CO, PAH and H{\sc i} maps. They manually selected spiral structures and investigated arm/interarm variations, showing the same clear peaks in the radial profiles for individual arms. Finally, \citet{2022ApJ...930..170H} built a decomposition model using {\small GALFIT}, and clearly demonstrated the mentioned gap in luminosity in their fig.~8.

\subsubsection{Characteristic widths}

Arms' characteristic widths are presented in Fig.~\ref{fig:width}. We use the FWHM of radial slice of the spiral arm model as a measure of width. One should not interpret $w_e^\text{in} + w_e^\text{out}$ in eq.~\ref{eq:I_bot} as a perpendicular width itself, even if the arms do not widen ($\xi = 0$). In the S\'ersic profile (eq.~\ref{eq:Sersic}), a radially symmetric 2D distribution of light is implied, and $r_\text{eff}$ is the radius enclosing half of light. The dependence between $b_n$ and $n$ is determined by this condition (\citealt{2005PASA...22..118G}). Eq.~\ref{eq:I_bot}, where $w_e$ is analogous to $r_\text{eff}$, replicates the form of the S\'ersic law, but S\'ersic-like distribution of light here is only one-dimensional. Therefore, $w_e$ cannot be considered as a half-light width, and we use the FWHM instead. This definition produces larger arm's width estimates than may be intuitively expected, in our case around 1.5--2~kpc. Fig.~\ref{fig:width} shows a similar behaviour of both the arms, i.e. with a global maximum in optical/NIR bands and a second local peak in FIR bands. 
This dependence could be explained by the fact that the filters indicating narrower arms are associated with ongoing star formation, which can not propagate far enough from the place of birth. This explanation is supported by \cite{2023A&A...673A.147P} where the authors note a clear increase in the width of the spiral arms with increasing age of the stellar populations. Other datasets used confirm these findings, but the images convolved to the 24~$\upmu$m resolution can expose only the first peak. It is also important to mention that the arms' characteristic widths for the original resolution are 20\% larger in the optical bands, as shown in Fig.~\ref{fig:width}.
\par
There are only few arm width measurements available in the literature, particularly for M\,51. \citet{1982ApJ...253..101K} measured the arm width in M\,51 using H{\sc ii} regions and estimated a perpendicular width equal to 17~arcsec. Also using H{\sc ii} data, \citet{2015ApJ...800...53H} fit log-periodic  models to segments of each arm in M\,51 and measure width values up to 1~kpc. At the same time, spirals selected instead using photometric bands like in \citet{2010ApJ...725..534F} look significantly wider than measured in \citet{2015ApJ...800...53H} and \citet{1982ApJ...253..101K}, validating our result. There are two reasons why we obtain higher width values. Firstly, pitch angle is small, then values in Fig.~\ref{fig:width} should be closer to perpendicular width, but slightly larger than it. Second, H{\sc ii} regions, used in previous studies \citep{1982ApJ...253..101K,2015ApJ...800...53H}, are associated with ongoing star formation of only few Myr old, which can not travel far enough from spiral. However, UV band associated with much older stars of approximately 100~Myr age, hence resulting in a larger extent and explain the discrepancy. This explanation is supported by the observed width drop in $8-24~\upmu$m, where hot dust demonstrate the presence of young stars of ages $\leq 10$~Myr.
\citet{savchenko} estimated average arm width to be $(0.16\pm0.04)\times r_{25}$ for grand design galaxies in the $r$ band, see their figure~16. Taking an M\,51 disk size to be 5.6~arcmin \citep{1995yCat.7155....0D}, the resultant average width from that formula will be approximately 50~arcsec, similar to what is obtained in this work if we assume that the measured width is close to the perpendicular one.
\par
We do not note a noticeable arm width increase in the models toward the galaxy edge, i.e. we find $\xi = 0$ for every image and arm. Similarly, \citet{2015ApJ...800...53H} found that the arm width increases towards the edge in M\,51, but only slowly, which is almost unnoticeable in their images. As found in \citet{savchenko}, this slow increase of spirals' widths is typical among their sample of 155 galaxies.

\subsection{Discussion and remarks}

Indeed, previous authors tried to implement various models of spiral arms. \citet{2010AJ....139.2097P} introduce a new version of {\small GALFIT} with various non-axisymmetric models including spirals modeled by coordinate rotation mode. Among other objects, that work also contains an implementation for an M\,51 model in figure~21, with a resultant $\chi^2=54.2$ in $r$ band. Despite the potential for fitting spirals with various forms, their model is more complex than used here; it has 103 free parameters in the best fit. Also, these parameters are difficult to interpret as it is, and their model could be difficult to implement. These are the reasons for why despite its power, such an approach is rarely used in practice. In one of such works, \citet{2014ApJ...780...69L} used that approach for spiral arm models in 4 galaxies, which are modeled separately from the disk following \citet{2010AJ....139.2097P}. They find examples of galaxies, where spiral arms, with or without other non-traditional components, are needed if one wants to describe the structure of bulges. In contrast, \citet{2017ApJ...845..114G} also used models similar to \citet{2010AJ....139.2097P} for decomposition in the $R$~band and obtained that spirals have a negligible effect on bulge parameters' estimation. Another common way to model spirals when working with radial profiles is to use Gaussians to fit ``bumps'' on profiles, produced by spiral arms, as \citet{2019ApJ...873...85D} and \citet{2011A&A...527A.109P} do. However, this approach is only feasible for a 1D decomposition, and should not be considered rigorous spiral arm modeling. We are also aware of several other published analytical models of spiral arms \citep{2002ApJS..142..261C,2010MNRAS.403.1625F}, which, to the best of our knowledge, were not used in the decomposition of real objects. 
\par

Of course, there is also another numerical approach to take spirals into account in the decomposition process. One could use synthetic images with artificial arms of various forms, often logarithmic, to check how they affect the extracted parameters. Such an approach was used by \cite{2022A&A...659A.141S,2020ApJ...900..178L,2012ApJS..199...33D} and others, and have the advantage of a priori knowledge about the components in the galaxy, but also bound to the exact form of the spirals model used. Since arms in real galaxies aren't smooth or symmetrical, and because we lack a broad piece of information on their observational properties, this strategy does not always produce correct results.
\par
There are also plenty of works where authors try to fit spirals by manually constructed shapes without any analytical model. Spirals could be selected by simple rules like choosing 45\% of the brightest pixels like  \citet{2010ApJ...725..534F} do. \citet{savchenko} traced spiral arms in 155 SDSS galaxies using perpendicular cuts fitted by a two-sided Gaussian function. \citet{2020ApJ...900..178L} and \cite{2021MNRAS.507.3923M} used so-called crowdsourced masks manually painted by volunteers. In \cite{2021A&A...647A.120B}, the authors used Deep Learning to train a model which marks a spiral pattern. All these approaches are very promising because they could fit spirals with complex forms, but to our knowledge, so far they were not used in decomposition studies and should suffer from human bias, thus should be difficult to reproduce.
\par
Besides accounting for the spiral arms in decomposition, another novelty of this work is the number of photometric bands we used for such a task, i.e. 17 bands from the UV part of the spectrum through to the FIR. The closest examples of such a multiband decomposition to our knowledge is the lenticular galaxy NGC\,3115, decomposed in \cite{2021MNRAS.504.2146B} using 11 bands from the UV to 4.6~$\upmu$m, and M\,81 (NGC\,3031) in \cite{2023ApJS..267...26G} where the authors use 20 bands from the UV to 5.8~$\upmu$m. Both of these studies, however, cover a smaller part of the spectrum and use only traditional components in decomposition. \cite{2023ApJS..267...26G} actually try to mask out spirals and estimate their influence on decomposition, and found little or no effect, which is curious considering how big the mask is, see their figure 15. In any case, this is a rare situation, usually only a few adjacent bands are utilized in decomposition. Most often these are all or some subset of the optical $ugriz$ filters for samples of various sizes \citep{2014ApJ...787...24B,2017A&A...598A..32M,2018MNRAS.473.4731K,2022MNRAS.516..942C}, up to millions of galaxies in \citet{2011ApJS..196...11S}. Several bands can also be used to make deep co-adds and then decomposed, as in \citet{2019MNRAS.486..390B}. One of the biggest up-to-date decomposition projects uses images from DESI Legacy \citep{2019AJ....157..168D} in the optical $grz$ bands and the probabilistic software \citep{2016ascl.soft04008L} to build models for all visible objects in $\approx40000$~square degrees of the sky. Such models are useful to discover faint components in galaxies \citep{2022MNRAS.512.1371M}. These models also reveals how often extended and bright spirals remain in residues, see e.g. predicted vote fractions in \citealt{2023MNRAS.526.4768W}. The amount of such cases directly highlight the importance of the presented work. 
\par
Outside of the optical part of the spectrum, only the NIR bands $JHK_S$ are used frequently 
\citep{2015MNRAS.454..806K,2020A&A...633A.104P,2021MNRAS.507.5952R}, while works where UV or MIR/FIR filters are used, such as \cite{2019A&A...622A.132M} for {\it Herschel} 100-500~$\upmu$m bands, are only occasional. It is important to mention an interesting approach to decompose several bands simultaneously, thus building a physically motivated and more robust fit, which was implemented in the {\small MegaMorph} package \citep{2014MNRAS.444.3603V}. In the series of works \citet{2013MNRAS.430..330H,2014MNRAS.441.1340V,2022MNRAS.513.2985R}, the authors apply this package to both optical and NIR bands, finding models of galaxies and their components using up to 9 filters simultaneously. 
\par


The lack of decomposition works, which cover images in the whole observed spectrum shows the difficulty to prepare such images, because a lot of untrivial processing and methodological issues should be solved. DustPedia project provides a unique opportunity for tasks such as the decomposition of M\,51 here, because most of the preprocessing issues are already successfully solved and the data is ready to use. The importance of covering as much of the spectrum as possible in decomposition studies arises from the possibility of fitting individual components' SED and thus recover formation histories (e.g. \citealp{2017ApJ...851...10E}) and physical properties for them. A successful example of an implementation of such an analysis is the previously mentioned galaxy, NGC\,3115 \citep{2021MNRAS.504.2146B}, whose SED was individually recovered for a bulge, two disks, and bar. Similar work was done for M\,81 separately for the bulge and the disk \citep{2023ApJS..267...26G}. Although the SED construction for different stellar populations \citep{2014A&A...571A..69D,radiativeM51} and morphological components in the axisymmetric case \citep{2023MNRAS.526..118I} was already solved with radiative transfer codes for M\,51, it is still vital for other galaxies. Moreover, we only start to understand the influence spiral arms have on construction of SEDs for classical components \citep{2011A&A...527A.109P}, hence a lot of future work awaits in this direction.

There are several limitations and flaws in our analysis. Firstly, real spiral arms are not log-normal, not smooth, and are contaminated by spurs, feathers, locally enhanced star formation areas, dust lanes, and tidal features. It is unclear if these features could deform arm models and whether they should be taken into account during decomposition or not. Surely, they could influence the decomposition by ``locking'' the fitting in a local minimum, but most of them are unlikely to affect the global parameters, such as its pitch angle or characteristic width. Second, the choice of the exact form of analytical expression to fit an arm is arbitrary. There is no clear theoretical reason why it should be a good approximation. However, in previous paper \citet{2024MNRAS.527.9605C} we use the same model for spiral arms with substantially different appearance in 29 galaxies and find good fitness between the model and data, and similar parameters' systematics of bulge and disk as here. This fact could also be the answer for the next objection that we don't know how typical M\,51 is, and hence whether we could generalize our obtained results or not. The mentioned work, first in this series, shed light to this question. Also, there could be potential limitations related to the fact that we used dust resolution for all images. Surely, it was shown many times (for example, \citealt{2021MNRAS.508.5825M}) that image resolution of the data and its PSF could affect the analysis. Indeed, we find that the adopted image quality of M\,51 largely affects the estimation of the S\'ersic parameter $n$ for the bulge, as seen in Fig.~\ref{fig:bulge}. However, for other parameters we find the different angular resolution and PSF to be less of an issue. Finally, we conclude that this work, while not free from limitations, produces values which are consistent with other works and with results for two auxiliary datasets, and we believe these facts give enough credibility to it.

\section{Conclusions}
\label{sec:conclusions}

Galaxies with a prominent spiral pattern are ubiquitous in the known Universe, but spiral arms are rarely taken into account in photometric decomposition of galaxies. We have done the work on this topic, and achieved the following results:


\par
(i) For the first time, we fitted decomposition models, consisting of a bulge, disk, and spiral arms, to images of M\,51 in 17 photometric bands from far UV to far IR. We found that models describe observations in all filters fairly well (Fig.~\ref{fig:main} and \ref{fig:slices}). We have shown that models with spiral arms (B+D+S) provide significantly smaller residuals than bulge+disk (B+D) models (Fig.~\ref{fig:bic} and \ref{fig:boxplots}), which justifies the large number of parameters of our spiral arm's model.
\par
(ii) We confirmed that spiral arms are the most luminous part of M\,51 in UV and MIR/FIR bands. We measured spiral arms' contribution to the total luminosity, which is roughly equal to 60\% in UV in FIR bands and $\geq30\%$ in other bands (Fig.~\ref{fig:frac}).
 \par
(iii) We have compared B+D+S and B+D models and found how neglecting the spiral arms affects the estimation of parameters. After including the spiral arms in the model, disk scale $h$ increases by 5--10\%, and its central intensity drops by a factor of 1.25--3 in most cases (Fig.~\ref{fig:disk}). For the bulge, which is small in M\,51 and severely affected by the image resolution used, obtained parameters should be taken with caution. WE find that separating the spiral arms into its own component allow one to resolve bulge properties in a larger number of bands. Thus, the S{\'e}rsic indices $n$ increase in all bands in B+D+S models, however, staying below unity in most of them and increasing with the used image resolution (Fig.~\ref{fig:bulge}). Bulge-to-total ratio increases in all bands. 
\par
(iv) In the best-fitting models with spiral arms, we derive arm pitch angles for the first time using a full 2D light distribution within the arms (Fig.~\ref{fig:pitch}). The derived pitch angle estimates for different bands agree within their uncertainties, in contrast to what would be expected if a long-standing density wave existed for the whole spiral pattern in M\,51. 
\par
(v) The arms' characteristic width in M\,51 is relatively constant with wavelength till the NIR, significantly drops after it and increases to nearly the same level for $\lambda \geq 70~\upmu$m  (Fig.~\ref{fig:width}). The radial scales of the arm intensity roughly follow the same distribution as the width, but for the arm pointed towards the NGC\,5195 it is significantly larger in most of the bands. The azimuthally-averaged spirals-to-total ratio demonstrates two peaks, expected from the literature, and contains from half to 80\% of emission in them (Fig.~\ref{fig:ST}).
\par

Mutatis mutandis, where possible our findings in this study confirm the results of \citet{2024MNRAS.527.9605C} and other authors. Our studies may be the first step in a new area of spiral arm analysis and will therefore be continued. In the next papers, we will conduct the similar analysis for more distant galaxies and continue to study variations in spiral structure at different wavelengths. Further improvement of the arm model is also a possible direction for future work. For example, a complex fit of bright star forming areas along the arm should be successfully modelled by the probabilistic approach used in DESI Legacy \citep{2019AJ....157..168D,2016ascl.soft04008L}.


\section*{Acknowledgements}
We acknowledge financial support from the Russian Science Foundation, grant no. 20-72-10052. 
\par
We thank the anonymous referee for his/her thorough review and highly appreciate the comments and suggestions that significantly contributed to improving the quality of the article. We thank Angelos Nersesian for providing errormaps for original photometric data. We thank Viviana Casasola for useful comments.
\par
DustPedia is a collaborative focused research project supported by the European Union under the Seventh Framework Programme (2007-2013) call (proposal no. 606847). The participating institutions are: Cardiff University, UK; National Observatory of Athens, Greece; Ghent University, Belgium; Université Paris Sud, France; National Institute for Astrophysics, Italy and CEA, France.

\section*{Data availability}
The data underlying this article will be shared on reasonable request to the corresponding author.
  
\bibliographystyle{mnras}
\bibliography{bibtex}

\begin{thebibliography}{}
\makeatletter
\relax
\def\mn@urlcharsother{\let\do\@makeother \do\$\do\&\do\#\do\^\do\_\do\%\do\~}
\def\mn@doi{\begingroup\mn@urlcharsother \@ifnextchar [ {\mn@doi@}
  {\mn@doi@[]}}
\def\mn@doi@[#1]#2{\def\@tempa{#1}\ifx\@tempa\@empty \href
  {http://dx.doi.org/#2} {doi:#2}\else \href {http://dx.doi.org/#2} {#1}\fi
  \endgroup}
\def\mn@eprint#1#2{\mn@eprint@#1:#2::\@nil}
\def\mn@eprint@arXiv#1{\href {http://arxiv.org/abs/#1} {{\tt arXiv:#1}}}
\def\mn@eprint@dblp#1{\href {http://dblp.uni-trier.de/rec/bibtex/#1.xml}
  {dblp:#1}}
\def\mn@eprint@#1:#2:#3:#4\@nil{\def\@tempa {#1}\def\@tempb {#2}\def\@tempc
  {#3}\ifx \@tempc \@empty \let \@tempc \@tempb \let \@tempb \@tempa \fi \ifx
  \@tempb \@empty \def\@tempb {arXiv}\fi \@ifundefined
  {mn@eprint@\@tempb}{\@tempb:\@tempc}{\expandafter \expandafter \csname
  mn@eprint@\@tempb\endcsname \expandafter{\@tempc}}}

\bibitem[\protect\citeauthoryear{{Abdeen}, {Kennefick}, {Kennefick}, {Miller},
  {Shields}, {Monson}  \& {Davis}}{{Abdeen} et~al.}{2020}]{2020MNRAS.496.1610A}
{Abdeen} S.,  {Kennefick} D.,  {Kennefick} J.,  {Miller} R.,  {Shields} D.~W.,
  {Monson} E.~B.,   {Davis} B.~L.,  2020, \mn@doi [\mnras]
  {10.1093/mnras/staa1596}, \href
  {https://ui.adsabs.harvard.edu/abs/2020MNRAS.496.1610A} {496, 1610}

\bibitem[\protect\citeauthoryear{{Aniano}, {Draine}, {Gordon}  \&
  {Sandstrom}}{{Aniano} et~al.}{2011}]{2011PASP..123.1218A}
{Aniano} G.,  {Draine} B.~T.,  {Gordon} K.~D.,   {Sandstrom} K.,  2011, \mn@doi
  [\pasp] {10.1086/662219}, \href
  {https://ui.adsabs.harvard.edu/abs/2011PASP..123.1218A} {123, 1218}

\bibitem[\protect\citeauthoryear{{Athanassoula}, {Laurikainen}, {Salo}  \&
  {Bosma}}{{Athanassoula} et~al.}{2015}]{2015MNRAS.454.3843A}
{Athanassoula} E.,  {Laurikainen} E.,  {Salo} H.,   {Bosma} A.,  2015, \mn@doi
  [\mnras] {10.1093/mnras/stv2231}, \href
  {https://ui.adsabs.harvard.edu/abs/2015MNRAS.454.3843A} {454, 3843}

\bibitem[\protect\citeauthoryear{{Baes} et~al.,}{{Baes}
  et~al.}{2020}]{2020A&A...641A.119B}
{Baes} M.,  et~al., 2020, \mn@doi [\aap] {10.1051/0004-6361/202038470}, \href
  {https://ui.adsabs.harvard.edu/abs/2020A&A...641A.119B} {641, A119}

\bibitem[\protect\citeauthoryear{{Baggett}, {Baggett}  \& {Anderson}}{{Baggett}
  et~al.}{1998}]{1998AJ....116.1626B}
{Baggett} W.~E.,  {Baggett} S.~M.,   {Anderson} K.~S.~J.,  1998, \mn@doi [\aj]
  {10.1086/300525}, \href
  {https://ui.adsabs.harvard.edu/abs/1998AJ....116.1626B} {116, 1626}

\bibitem[\protect\citeauthoryear{{Bailer-Jones}}{{Bailer-Jones}}{2017}]{2017pbi..book.....B}
{Bailer-Jones} C. A.~L.,  2017, {Practical Bayesian Inference}

\bibitem[\protect\citeauthoryear{{Barway}, {Wadadekar}, {Kembhavi}  \&
  {Mayya}}{{Barway} et~al.}{2009}]{2009MNRAS.394.1991B}
{Barway} S.,  {Wadadekar} Y.,  {Kembhavi} A.~K.,   {Mayya} Y.~D.,  2009,
  \mn@doi [\mnras] {10.1111/j.1365-2966.2009.14440.x}, \href
  {https://ui.adsabs.harvard.edu/abs/2009MNRAS.394.1991B} {394, 1991}

\bibitem[\protect\citeauthoryear{{Bekki}}{{Bekki}}{2021}]{2021A&A...647A.120B}
{Bekki} K.,  2021, \mn@doi [\aap] {10.1051/0004-6361/202039797}, \href
  {https://ui.adsabs.harvard.edu/abs/2021A&A...647A.120B} {647, A120}

\bibitem[\protect\citeauthoryear{{Bianchi} et~al.,}{{Bianchi}
  et~al.}{2018}]{2018A&A...620A.112B}
{Bianchi} S.,  et~al., 2018, \mn@doi [\aap] {10.1051/0004-6361/201833699},
  \href {https://ui.adsabs.harvard.edu/abs/2018A&A...620A.112B} {620, A112}

\bibitem[\protect\citeauthoryear{{Bizyaev}, {Kautsch}, {Mosenkov},
  {Reshetnikov}, {Sotnikova}, {Yablokova}  \& {Hillyer}}{{Bizyaev}
  et~al.}{2014}]{2014ApJ...787...24B}
{Bizyaev} D.~V.,  {Kautsch} S.~J.,  {Mosenkov} A.~V.,  {Reshetnikov} V.~P.,
  {Sotnikova} N.~Y.,  {Yablokova} N.~V.,   {Hillyer} R.~W.,  2014, \mn@doi
  [\apj] {10.1088/0004-637X/787/1/24}, \href
  {https://ui.adsabs.harvard.edu/abs/2014ApJ...787...24B} {787, 24}

\bibitem[\protect\citeauthoryear{{Bottrell}, {Simard}, {Mendel}  \&
  {Ellison}}{{Bottrell} et~al.}{2019}]{2019MNRAS.486..390B}
{Bottrell} C.,  {Simard} L.,  {Mendel} J.~T.,   {Ellison} S.~L.,  2019, \mn@doi
  [\mnras] {10.1093/mnras/stz855}, \href
  {https://ui.adsabs.harvard.edu/abs/2019MNRAS.486..390B} {486, 390}

\bibitem[\protect\citeauthoryear{{Bradley} et~al.,}{{Bradley}
  et~al.}{2020}]{2020zndo...4044744B}
{Bradley} L.,  et~al., 2020, {astropy/photutils: 1.0.0}, Zenodo,
  \mn@doi{10.5281/zenodo.4044744}

\bibitem[\protect\citeauthoryear{{Buzzo} et~al.,}{{Buzzo}
  et~al.}{2021}]{2021MNRAS.504.2146B}
{Buzzo} M.~L.,  et~al., 2021, \mn@doi [\mnras] {10.1093/mnras/stab941}, \href
  {https://ui.adsabs.harvard.edu/abs/2021MNRAS.504.2146B} {504, 2146}

\bibitem[\protect\citeauthoryear{{Casasola} et~al.,}{{Casasola}
  et~al.}{2017}]{2017A&A...605A..18C}
{Casasola} V.,  et~al., 2017, \mn@doi [\aap] {10.1051/0004-6361/201731020},
  \href {https://ui.adsabs.harvard.edu/abs/2017A&A...605A..18C} {605, A18}

\bibitem[\protect\citeauthoryear{{Casura} et~al.,}{{Casura}
  et~al.}{2022}]{2022MNRAS.516..942C}
{Casura} S.,  et~al., 2022, \mn@doi [\mnras] {10.1093/mnras/stac2267}, \href
  {https://ui.adsabs.harvard.edu/abs/2022MNRAS.516..942C} {516, 942}

\bibitem[\protect\citeauthoryear{{Chandar} et~al.,}{{Chandar}
  et~al.}{2017}]{2017ApJ...845...78C}
{Chandar} R.,  et~al., 2017, \mn@doi [\apj] {10.3847/1538-4357/aa7b38}, \href
  {https://ui.adsabs.harvard.edu/abs/2017ApJ...845...78C} {845, 78}

\bibitem[\protect\citeauthoryear{{Chugunov} et~al.,}{{Chugunov}
  et~al.}{2024}]{2024MNRAS.527.9605C}
{Chugunov} I.~V.,  et~al., 2024, \mn@doi [\mnras] {10.1093/mnras/stad3850},
  \href {https://ui.adsabs.harvard.edu/abs/2024MNRAS.527.9605C} {527, 9605}

\bibitem[\protect\citeauthoryear{{Clark} et~al.,}{{Clark}
  et~al.}{2018}]{dustpedia}
{Clark} C.~J.~R.,  et~al., 2018, \mn@doi [\aap] {10.1051/0004-6361/201731419},
  \href {https://ui.adsabs.harvard.edu/abs/2018A&A...609A..37C} {609, A37}

\bibitem[\protect\citeauthoryear{{Clark} et~al.,}{{Clark}
  et~al.}{2019}]{2019MNRAS.489.5256C}
{Clark} C.~J.~R.,  et~al., 2019, \mn@doi [\mnras] {10.1093/mnras/stz2257},
  \href {https://ui.adsabs.harvard.edu/abs/2019MNRAS.489.5256C} {489, 5256}

\bibitem[\protect\citeauthoryear{{Conselice}}{{Conselice}}{1997}]{1997PASP..109.1251C}
{Conselice} C.~J.,  1997, \mn@doi [\pasp] {10.1086/134004}, \href
  {https://ui.adsabs.harvard.edu/abs/1997PASP..109.1251C} {109, 1251}

\bibitem[\protect\citeauthoryear{{Conselice}}{{Conselice}}{2006}]{2006MNRAS.373.1389C}
{Conselice} C.~J.,  2006, \mn@doi [\mnras] {10.1111/j.1365-2966.2006.11114.x},
  \href {https://ui.adsabs.harvard.edu/abs/2006MNRAS.373.1389C} {373, 1389}

\bibitem[\protect\citeauthoryear{{Cox} \& {G{\'o}mez}}{{Cox} \&
  {G{\'o}mez}}{2002}]{2002ApJS..142..261C}
{Cox} D.~P.,  {G{\'o}mez} G.~C.,  2002, \mn@doi [\apjs] {10.1086/341946}, \href
  {https://ui.adsabs.harvard.edu/abs/2002ApJS..142..261C} {142, 261}

\bibitem[\protect\citeauthoryear{{D'Souza}, {Kauffman}, {Wang}  \&
  {Vegetti}}{{D'Souza} et~al.}{2014}]{2014MNRAS.443.1433D}
{D'Souza} R.,  {Kauffman} G.,  {Wang} J.,   {Vegetti} S.,  2014, \mn@doi
  [\mnras] {10.1093/mnras/stu1194}, \href
  {https://ui.adsabs.harvard.edu/abs/2014MNRAS.443.1433D} {443, 1433}

\bibitem[\protect\citeauthoryear{{Daigle}, {Carignan}, {Amram}, {Hernandez},
  {Chemin}, {Balkowski}  \& {Kennicutt}}{{Daigle}
  et~al.}{2006}]{2006MNRAS.367..469D}
{Daigle} O.,  {Carignan} C.,  {Amram} P.,  {Hernandez} O.,  {Chemin} L.,
  {Balkowski} C.,   {Kennicutt} R.,  2006, \mn@doi [\mnras]
  {10.1111/j.1365-2966.2006.10002.x}, \href
  {https://ui.adsabs.harvard.edu/abs/2006MNRAS.367..469D} {367, 469}

\bibitem[\protect\citeauthoryear{{Davis}, {Berrier}, {Shields}, {Kennefick},
  {Kennefick}, {Seigar}, {Lacy}  \& {Puerari}}{{Davis}
  et~al.}{2012}]{2012ApJS..199...33D}
{Davis} B.~L.,  {Berrier} J.~C.,  {Shields} D.~W.,  {Kennefick} J.,
  {Kennefick} D.,  {Seigar} M.~S.,  {Lacy} C. H.~S.,   {Puerari} I.,  2012,
  \mn@doi [\apjs] {10.1088/0067-0049/199/2/33}, \href
  {https://ui.adsabs.harvard.edu/abs/2012ApJS..199...33D} {199, 33}

\bibitem[\protect\citeauthoryear{{Davis}, {Graham}  \& {Cameron}}{{Davis}
  et~al.}{2019}]{2019ApJ...873...85D}
{Davis} B.~L.,  {Graham} A.~W.,   {Cameron} E.,  2019, \mn@doi [\apj]
  {10.3847/1538-4357/aaf3b8}, \href
  {https://ui.adsabs.harvard.edu/abs/2019ApJ...873...85D} {873, 85}

\bibitem[\protect\citeauthoryear{{De Looze} et~al.,}{{De Looze}
  et~al.}{2014}]{2014A&A...571A..69D}
{De Looze} I.,  et~al., 2014, \mn@doi [\aap] {10.1051/0004-6361/201424747},
  \href {https://ui.adsabs.harvard.edu/abs/2014A&A...571A..69D} {571, A69}

\bibitem[\protect\citeauthoryear{{De Vis} et~al.,}{{De Vis}
  et~al.}{2019}]{2019A&A...623A...5D}
{De Vis} P.,  et~al., 2019, \mn@doi [\aap] {10.1051/0004-6361/201834444}, \href
  {https://ui.adsabs.harvard.edu/abs/2019A&A...623A...5D} {623, A5}

\bibitem[\protect\citeauthoryear{{Dessart} et~al.,}{{Dessart}
  et~al.}{2008}]{2008ApJ...675..644D}
{Dessart} L.,  et~al., 2008, \mn@doi [\apj] {10.1086/526451}, \href
  {https://ui.adsabs.harvard.edu/abs/2008ApJ...675..644D} {675, 644}

\bibitem[\protect\citeauthoryear{{Dey} et~al.,}{{Dey}
  et~al.}{2019}]{2019AJ....157..168D}
{Dey} A.,  et~al., 2019, \mn@doi [\aj] {10.3847/1538-3881/ab089d}, \href
  {https://ui.adsabs.harvard.edu/abs/2019AJ....157..168D} {157, 168}

\bibitem[\protect\citeauthoryear{{D{\'\i}az-Garc{\'\i}a}, {Salo}, {Knapen}  \&
  {Herrera-Endoqui}}{{D{\'\i}az-Garc{\'\i}a}
  et~al.}{2019}]{2019A&A...631A..94D}
{D{\'\i}az-Garc{\'\i}a} S.,  {Salo} H.,  {Knapen} J.~H.,   {Herrera-Endoqui}
  M.,  2019, \mn@doi [\aap] {10.1051/0004-6361/201936000}, \href
  {https://ui.adsabs.harvard.edu/abs/2019A&A...631A..94D} {631, A94}

\bibitem[\protect\citeauthoryear{{Dobbs}, {Theis}, {Pringle}  \&
  {Bate}}{{Dobbs} et~al.}{2010}]{2010MNRAS.403..625D}
{Dobbs} C.~L.,  {Theis} C.,  {Pringle} J.~E.,   {Bate} M.~R.,  2010, \mn@doi
  [\mnras] {10.1111/j.1365-2966.2009.16161.x}, \href
  {https://ui.adsabs.harvard.edu/abs/2010MNRAS.403..625D} {403, 625}

\bibitem[\protect\citeauthoryear{{Egusa}, {Kohno}, {Sofue}, {Nakanishi}  \&
  {Komugi}}{{Egusa} et~al.}{2009}]{2009ApJ...697.1870E}
{Egusa} F.,  {Kohno} K.,  {Sofue} Y.,  {Nakanishi} H.,   {Komugi} S.,  2009,
  \mn@doi [\apj] {10.1088/0004-637X/697/2/1870}, \href
  {https://ui.adsabs.harvard.edu/abs/2009ApJ...697.1870E} {697, 1870}

\bibitem[\protect\citeauthoryear{{Elmegreen}}{{Elmegreen}}{1990}]{1990NYASA.596...40E}
{Elmegreen} B.~G.,  1990, \mn@doi [Annals of the New York Academy of Sciences]
  {10.1111/j.1749-6632.1990.tb27410.x}, \href
  {https://ui.adsabs.harvard.edu/abs/1990NYASA.596...40E} {596, 40}

\bibitem[\protect\citeauthoryear{{Elmegreen} \& {Elmegreen}}{{Elmegreen} \&
  {Elmegreen}}{1995}]{1995ApJ...445..591E}
{Elmegreen} D.~M.,  {Elmegreen} B.~G.,  1995, \mn@doi [\apj] {10.1086/175723},
  \href {https://ui.adsabs.harvard.edu/abs/1995ApJ...445..591E} {445, 591}

\bibitem[\protect\citeauthoryear{{Elmegreen}, {Elmegreen}  \&
  {Seiden}}{{Elmegreen} et~al.}{1989}]{1989ApJ...343..602E}
{Elmegreen} B.~G.,  {Elmegreen} D.~M.,   {Seiden} P.~E.,  1989, \mn@doi [\apj]
  {10.1086/167733}, \href
  {https://ui.adsabs.harvard.edu/abs/1989ApJ...343..602E} {343, 602}

\bibitem[\protect\citeauthoryear{{Erwin}}{{Erwin}}{2015}]{2015ApJ...799..226E}
{Erwin} P.,  2015, \mn@doi [\apj] {10.1088/0004-637X/799/2/226}, \href
  {https://ui.adsabs.harvard.edu/abs/2015ApJ...799..226E} {799, 226}

\bibitem[\protect\citeauthoryear{{Eufrasio} et~al.,}{{Eufrasio}
  et~al.}{2017}]{2017ApJ...851...10E}
{Eufrasio} R.~T.,  et~al., 2017, \mn@doi [\apj] {10.3847/1538-4357/aa9569},
  \href {https://ui.adsabs.harvard.edu/abs/2017ApJ...851...10E} {851, 10}

\bibitem[\protect\citeauthoryear{{Fisher} \& {Drory}}{{Fisher} \&
  {Drory}}{2008}]{2008AJ....136..773F}
{Fisher} D.~B.,  {Drory} N.,  2008, \mn@doi [\aj]
  {10.1088/0004-6256/136/2/773}, \href
  {https://ui.adsabs.harvard.edu/abs/2008AJ....136..773F} {136, 773}

\bibitem[\protect\citeauthoryear{{Foyle}, {Rix}, {Walter}  \& {Leroy}}{{Foyle}
  et~al.}{2010}]{2010ApJ...725..534F}
{Foyle} K.,  {Rix} H.~W.,  {Walter} F.,   {Leroy} A.~K.,  2010, \mn@doi [\apj]
  {10.1088/0004-637X/725/1/534}, \href
  {https://ui.adsabs.harvard.edu/abs/2010ApJ...725..534F} {725, 534}

\bibitem[\protect\citeauthoryear{{Fridman} \& {Poltorak}}{{Fridman} \&
  {Poltorak}}{2010}]{2010MNRAS.403.1625F}
{Fridman} A.~M.,  {Poltorak} S.~G.,  2010, \mn@doi [\mnras]
  {10.1111/j.1365-2966.2010.16229.x}, \href
  {https://ui.adsabs.harvard.edu/abs/2010MNRAS.403.1625F} {403, 1625}

\bibitem[\protect\citeauthoryear{{Gadotti}}{{Gadotti}}{2008}]{2008MNRAS.384..420G}
{Gadotti} D.~A.,  2008, \mn@doi [\mnras] {10.1111/j.1365-2966.2007.12723.x},
  \href {https://ui.adsabs.harvard.edu/abs/2008MNRAS.384..420G} {384, 420}

\bibitem[\protect\citeauthoryear{{Gadotti}}{{Gadotti}}{2009}]{2009MNRAS.393.1531G}
{Gadotti} D.~A.,  2009, \mn@doi [\mnras] {10.1111/j.1365-2966.2008.14257.x},
  \href {https://ui.adsabs.harvard.edu/abs/2009MNRAS.393.1531G} {393, 1531}

\bibitem[\protect\citeauthoryear{{Gadotti} et~al.,}{{Gadotti}
  et~al.}{2020}]{2020A&A...643A..14G}
{Gadotti} D.~A.,  et~al., 2020, \mn@doi [\aap] {10.1051/0004-6361/202038448},
  \href {https://ui.adsabs.harvard.edu/abs/2020A&A...643A..14G} {643, A14}

\bibitem[\protect\citeauthoryear{{Gao} \& {Ho}}{{Gao} \&
  {Ho}}{2017}]{2017ApJ...845..114G}
{Gao} H.,  {Ho} L.~C.,  2017, \mn@doi [\apj] {10.3847/1538-4357/aa7da4}, \href
  {https://ui.adsabs.harvard.edu/abs/2017ApJ...845..114G} {845, 114}

\bibitem[\protect\citeauthoryear{{Gong}, {Mao}, {Gao}  \& {Yu}}{{Gong}
  et~al.}{2023}]{2023ApJS..267...26G}
{Gong} J.-Y.,  {Mao} Y.-W.,  {Gao} H.,   {Yu} S.-Y.,  2023, \mn@doi [\apjs]
  {10.3847/1538-4365/acd554}, \href
  {https://ui.adsabs.harvard.edu/abs/2023ApJS..267...26G} {267, 26}

\bibitem[\protect\citeauthoryear{{Graham} \& {Driver}}{{Graham} \&
  {Driver}}{2005}]{2005PASA...22..118G}
{Graham} A.~W.,  {Driver} S.~P.,  2005, \mn@doi [\pasa] {10.1071/AS05001},
  \href {https://ui.adsabs.harvard.edu/abs/2005PASA...22..118G} {22, 118}

\bibitem[\protect\citeauthoryear{{H{\"a}u{\ss}ler} et~al.,}{{H{\"a}u{\ss}ler}
  et~al.}{2013}]{2013MNRAS.430..330H}
{H{\"a}u{\ss}ler} B.,  et~al., 2013, \mn@doi [\mnras] {10.1093/mnras/sts633},
  \href {https://ui.adsabs.harvard.edu/abs/2013MNRAS.430..330H} {430, 330}

\bibitem[\protect\citeauthoryear{{Head}, {Lucey}, {Hudson}  \& {Smith}}{{Head}
  et~al.}{2014}]{2014MNRAS.440.1690H}
{Head} J. T.~C.~G.,  {Lucey} J.~R.,  {Hudson} M.~J.,   {Smith} R.~J.,  2014,
  \mn@doi [\mnras] {10.1093/mnras/stu325}, \href
  {https://ui.adsabs.harvard.edu/abs/2014MNRAS.440.1690H} {440, 1690}

\bibitem[\protect\citeauthoryear{{Heyer} et~al.,}{{Heyer}
  et~al.}{2022}]{2022ApJ...930..170H}
{Heyer} M.,  et~al., 2022, \mn@doi [\apj] {10.3847/1538-4357/ac67ea}, \href
  {https://ui.adsabs.harvard.edu/abs/2022ApJ...930..170H} {930, 170}

\bibitem[\protect\citeauthoryear{{Honig} \& {Reid}}{{Honig} \&
  {Reid}}{2015}]{2015ApJ...800...53H}
{Honig} Z.~N.,  {Reid} M.~J.,  2015, \mn@doi [\apj]
  {10.1088/0004-637X/800/1/53}, \href
  {https://ui.adsabs.harvard.edu/abs/2015ApJ...800...53H} {800, 53}

\bibitem[\protect\citeauthoryear{{Hu}, {Shao}  \& {Peng}}{{Hu}
  et~al.}{2013}]{2013ApJ...762L..27H}
{Hu} T.,  {Shao} Z.,   {Peng} Q.,  2013, \mn@doi [\apjl]
  {10.1088/2041-8205/762/2/L27}, \href
  {https://ui.adsabs.harvard.edu/abs/2013ApJ...762L..27H} {762, L27}

\bibitem[\protect\citeauthoryear{{Inman}, {Popescu}, {Rushton}  \&
  {Murphy}}{{Inman} et~al.}{2023}]{2023MNRAS.526..118I}
{Inman} C.~J.,  {Popescu} C.~C.,  {Rushton} M.~T.,   {Murphy} D.,  2023,
  \mn@doi [\mnras] {10.1093/mnras/stad2676}, \href
  {https://ui.adsabs.harvard.edu/abs/2023MNRAS.526..118I} {526, 118}

\bibitem[\protect\citeauthoryear{{Kelvin} et~al.,}{{Kelvin}
  et~al.}{2012}]{2012MNRAS.421.1007K}
{Kelvin} L.~S.,  et~al., 2012, \mn@doi [\mnras]
  {10.1111/j.1365-2966.2012.20355.x}, \href
  {https://ui.adsabs.harvard.edu/abs/2012MNRAS.421.1007K} {421, 1007}

\bibitem[\protect\citeauthoryear{{Kendall}, {Kennicutt}  \& {Clarke}}{{Kendall}
  et~al.}{2011}]{2011MNRAS.414..538K}
{Kendall} S.,  {Kennicutt} R.~C.,   {Clarke} C.,  2011, \mn@doi [\mnras]
  {10.1111/j.1365-2966.2011.18422.x}, \href
  {https://ui.adsabs.harvard.edu/abs/2011MNRAS.414..538K} {414, 538}

\bibitem[\protect\citeauthoryear{{Kendall}, {Clarke}  \& {Kennicutt}}{{Kendall}
  et~al.}{2015}]{2015MNRAS.446.4155K}
{Kendall} S.,  {Clarke} C.,   {Kennicutt} R.~C.,  2015, \mn@doi [\mnras]
  {10.1093/mnras/stu2431}, \href
  {https://ui.adsabs.harvard.edu/abs/2015MNRAS.446.4155K} {446, 4155}

\bibitem[\protect\citeauthoryear{{Kennedy} et~al.,}{{Kennedy}
  et~al.}{2015}]{2015MNRAS.454..806K}
{Kennedy} R.,  et~al., 2015, \mn@doi [\mnras] {10.1093/mnras/stv2032}, \href
  {https://ui.adsabs.harvard.edu/abs/2015MNRAS.454..806K} {454, 806}

\bibitem[\protect\citeauthoryear{{Kennicutt}}{{Kennicutt}}{1981}]{1981AJ.....86.1847K}
{Kennicutt} R.~C. J.,  1981, \mn@doi [\aj] {10.1086/113064}, \href
  {https://ui.adsabs.harvard.edu/abs/1981AJ.....86.1847K} {86, 1847}

\bibitem[\protect\citeauthoryear{{Kennicutt} \& {Hodge}}{{Kennicutt} \&
  {Hodge}}{1982}]{1982ApJ...253..101K}
{Kennicutt} R. J.,  {Hodge} P.,  1982, \mn@doi [\apj] {10.1086/159614}, \href
  {https://ui.adsabs.harvard.edu/abs/1982ApJ...253..101K} {253, 101}

\bibitem[\protect\citeauthoryear{{Knapen}, {Beckman}, {Cepa}, {van der Hulst}
  \& {Rand}}{{Knapen} et~al.}{1992}]{1992ApJ...385L..37K}
{Knapen} J.~H.,  {Beckman} J.~E.,  {Cepa} J.,  {van der Hulst} T.,   {Rand}
  R.~J.,  1992, \mn@doi [\apjl] {10.1086/186272}, \href
  {https://ui.adsabs.harvard.edu/abs/1992ApJ...385L..37K} {385, L37}

\bibitem[\protect\citeauthoryear{{Kruk} et~al.,}{{Kruk}
  et~al.}{2018}]{2018MNRAS.473.4731K}
{Kruk} S.~J.,  et~al., 2018, \mn@doi [\mnras] {10.1093/mnras/stx2605}, \href
  {https://ui.adsabs.harvard.edu/abs/2018MNRAS.473.4731K} {473, 4731}

\bibitem[\protect\citeauthoryear{{La Barbera}, {de Carvalho}, {de La Rosa},
  {Lopes}, {Kohl-Moreira}  \& {Capelato}}{{La Barbera}
  et~al.}{2010}]{2010MNRAS.408.1313L}
{La Barbera} F.,  {de Carvalho} R.~R.,  {de La Rosa} I.~G.,  {Lopes} P.~A.~A.,
  {Kohl-Moreira} J.~L.,   {Capelato} H.~V.,  2010, \mn@doi [\mnras]
  {10.1111/j.1365-2966.2010.16850.x}, \href
  {https://ui.adsabs.harvard.edu/abs/2010MNRAS.408.1313L} {408, 1313}

\bibitem[\protect\citeauthoryear{{La Vigne}, {Vogel}  \& {Ostriker}}{{La Vigne}
  et~al.}{2006}]{2006ApJ...650..818L}
{La Vigne} M.~A.,  {Vogel} S.~N.,   {Ostriker} E.~C.,  2006, \mn@doi [\apj]
  {10.1086/506589}, \href
  {https://ui.adsabs.harvard.edu/abs/2006ApJ...650..818L} {650, 818}

\bibitem[\protect\citeauthoryear{{Laine} et~al.,}{{Laine}
  et~al.}{2014}]{2014MNRAS.441.1992L}
{Laine} J.,  et~al., 2014, \mn@doi [\mnras] {10.1093/mnras/stu628}, \href
  {https://ui.adsabs.harvard.edu/abs/2014MNRAS.441.1992L} {441, 1992}

\bibitem[\protect\citeauthoryear{{Lang}, {Hogg}  \& {Mykytyn}}{{Lang}
  et~al.}{2016}]{2016ascl.soft04008L}
{Lang} D.,  {Hogg} D.~W.,   {Mykytyn} D.,  2016, {The Tractor: Probabilistic
  astronomical source detection and measurement}, Astrophysics Source Code
  Library, record ascl:1604.008 (\mn@eprint {ascl} {1604.008})

\bibitem[\protect\citeauthoryear{{L{\"a}sker}, {Ferrarese}  \& {van de
  Ven}}{{L{\"a}sker} et~al.}{2014}]{2014ApJ...780...69L}
{L{\"a}sker} R.,  {Ferrarese} L.,   {van de Ven} G.,  2014, \mn@doi [\apj]
  {10.1088/0004-637X/780/1/69}, \href
  {https://ui.adsabs.harvard.edu/abs/2014ApJ...780...69L} {780, 69}

\bibitem[\protect\citeauthoryear{{Laurikainen} \& {Salo}}{{Laurikainen} \&
  {Salo}}{2017}]{2017A&A...598A..10L}
{Laurikainen} E.,  {Salo} H.,  2017, \mn@doi [\aap]
  {10.1051/0004-6361/201628936}, \href
  {https://ui.adsabs.harvard.edu/abs/2017A&A...598A..10L} {598, A10}

\bibitem[\protect\citeauthoryear{{Laurikainen}, {Salo}, {Buta}, {Knapen},
  {Speltincx}  \& {Block}}{{Laurikainen} et~al.}{2006}]{2006AJ....132.2634L}
{Laurikainen} E.,  {Salo} H.,  {Buta} R.,  {Knapen} J.,  {Speltincx} T.,
  {Block} D.,  2006, \mn@doi [\aj] {10.1086/508810}, \href
  {https://ui.adsabs.harvard.edu/abs/2006AJ....132.2634L} {132, 2634}

\bibitem[\protect\citeauthoryear{{Laurikainen}, {Salo}, {Buta}  \&
  {Knapen}}{{Laurikainen} et~al.}{2007}]{2007MNRAS.381..401L}
{Laurikainen} E.,  {Salo} H.,  {Buta} R.,   {Knapen} J.~H.,  2007, \mn@doi
  [\mnras] {10.1111/j.1365-2966.2007.12299.x}, \href
  {https://ui.adsabs.harvard.edu/abs/2007MNRAS.381..401L} {381, 401}

\bibitem[\protect\citeauthoryear{{Laurikainen}, {Salo}, {Athanassoula}, {Bosma}
   \& {Herrera-Endoqui}}{{Laurikainen} et~al.}{2014}]{2014MNRAS.444L..80L}
{Laurikainen} E.,  {Salo} H.,  {Athanassoula} E.,  {Bosma} A.,
  {Herrera-Endoqui} M.,  2014, \mn@doi [\mnras] {10.1093/mnrasl/slu118}, \href
  {https://ui.adsabs.harvard.edu/abs/2014MNRAS.444L..80L} {444, L80}

\bibitem[\protect\citeauthoryear{{Lingard} et~al.,}{{Lingard}
  et~al.}{2020}]{2020ApJ...900..178L}
{Lingard} T.~K.,  et~al., 2020, \mn@doi [\apj] {10.3847/1538-4357/ab9d83},
  \href {https://ui.adsabs.harvard.edu/abs/2020ApJ...900..178L} {900, 178}

\bibitem[\protect\citeauthoryear{{Ma}}{{Ma}}{2001}]{2001ChJAA...1..395M}
{Ma} J.,  2001, \mn@doi [\cjaa] {10.1088/1009-9271/1/5/395}, \href
  {https://ui.adsabs.harvard.edu/abs/2001ChJAA...1..395M} {1, 395}

\bibitem[\protect\citeauthoryear{{Makarov} et~al.,}{{Makarov}
  et~al.}{2022}]{2022MNRAS.511.3063M}
{Makarov} D.,  et~al., 2022, \mn@doi [\mnras] {10.1093/mnras/stac227}, \href
  {https://ui.adsabs.harvard.edu/abs/2022MNRAS.511.3063M} {511, 3063}

\bibitem[\protect\citeauthoryear{{Marchuk}, {Smirnov}, {Mosenkov}, {Il'in},
  {Gontcharov}, {Savchenko}  \& {Rom{\'a}n}}{{Marchuk}
  et~al.}{2021}]{2021MNRAS.508.5825M}
{Marchuk} A.~A.,  {Smirnov} A.~A.,  {Mosenkov} A.~V.,  {Il'in} V.~B.,
  {Gontcharov} G.~A.,  {Savchenko} S.~S.,   {Rom{\'a}n} J.,  2021, \mn@doi
  [\mnras] {10.1093/mnras/stab2846}, \href
  {https://ui.adsabs.harvard.edu/abs/2021MNRAS.508.5825M} {508, 5825}

\bibitem[\protect\citeauthoryear{{Marchuk} et~al.,}{{Marchuk}
  et~al.}{2022}]{2022MNRAS.512.1371M}
{Marchuk} A.~A.,  et~al., 2022, \mn@doi [\mnras] {10.1093/mnras/stac599}, \href
  {https://ui.adsabs.harvard.edu/abs/2022MNRAS.512.1371M} {512, 1371}

\bibitem[\protect\citeauthoryear{{Marinova} \& {Jogee}}{{Marinova} \&
  {Jogee}}{2007}]{2007ApJ...659.1176M}
{Marinova} I.,  {Jogee} S.,  2007, \mn@doi [\apj] {10.1086/512355}, \href
  {https://ui.adsabs.harvard.edu/abs/2007ApJ...659.1176M} {659, 1176}

\bibitem[\protect\citeauthoryear{{Mart{\'\i}nez-Garc{\'\i}a},
  {Gonz{\'a}lez-L{\'o}pezlira}  \& {Puerari}}{{Mart{\'\i}nez-Garc{\'\i}a}
  et~al.}{2023}]{2023MNRAS.524...18M}
{Mart{\'\i}nez-Garc{\'\i}a} E.~E.,  {Gonz{\'a}lez-L{\'o}pezlira} R.~A.,
  {Puerari} I.,  2023, \mn@doi [\mnras] {10.1093/mnras/stad1805}, \href
  {https://ui.adsabs.harvard.edu/abs/2023MNRAS.524...18M} {524, 18}

\bibitem[\protect\citeauthoryear{{Masters} et~al.,}{{Masters}
  et~al.}{2011}]{2011MNRAS.411.2026M}
{Masters} K.~L.,  et~al., 2011, \mn@doi [\mnras]
  {10.1111/j.1365-2966.2010.17834.x}, \href
  {https://ui.adsabs.harvard.edu/abs/2011MNRAS.411.2026M} {411, 2026}

\bibitem[\protect\citeauthoryear{{Masters} et~al.,}{{Masters}
  et~al.}{2021}]{2021MNRAS.507.3923M}
{Masters} K.~L.,  et~al., 2021, \mn@doi [\mnras] {10.1093/mnras/stab2282},
  \href {https://ui.adsabs.harvard.edu/abs/2021MNRAS.507.3923M} {507, 3923}

\bibitem[\protect\citeauthoryear{{McCall}}{{McCall}}{1982}]{1982PhDT........35M}
{McCall} M.~L.,  1982, PhD thesis, University of Texas, Austin

\bibitem[\protect\citeauthoryear{{Meidt}, {Rand}, {Merrifield}, {Shetty}  \&
  {Vogel}}{{Meidt} et~al.}{2008}]{2008ApJ...688..224M}
{Meidt} S.~E.,  {Rand} R.~J.,  {Merrifield} M.~R.,  {Shetty} R.,   {Vogel}
  S.~N.,  2008, \mn@doi [\apj] {10.1086/591516}, \href
  {https://ui.adsabs.harvard.edu/abs/2008ApJ...688..224M} {688, 224}

\bibitem[\protect\citeauthoryear{{M{\'e}ndez-Abreu} et~al.,}{{M{\'e}ndez-Abreu}
  et~al.}{2017}]{2017A&A...598A..32M}
{M{\'e}ndez-Abreu} J.,  et~al., 2017, \mn@doi [\aap]
  {10.1051/0004-6361/201629525}, \href
  {https://ui.adsabs.harvard.edu/abs/2017A&A...598A..32M} {598, A32}

\bibitem[\protect\citeauthoryear{{Miller}, {Kennefick}, {Kennefick}, {Shameer
  Abdeen}, {Monson}, {Eufrasio}, {Shields}  \& {Davis}}{{Miller}
  et~al.}{2019}]{2019ApJ...874..177M}
{Miller} R.,  {Kennefick} D.,  {Kennefick} J.,  {Shameer Abdeen} M.,  {Monson}
  E.,  {Eufrasio} R.~T.,  {Shields} D.~W.,   {Davis} B.~L.,  2019, \mn@doi
  [\apj] {10.3847/1538-4357/ab0d26}, \href
  {https://ui.adsabs.harvard.edu/abs/2019ApJ...874..177M} {874, 177}

\bibitem[\protect\citeauthoryear{{Morrissey} et~al.,}{{Morrissey}
  et~al.}{2007}]{2007ApJS..173..682M}
{Morrissey} P.,  et~al., 2007, \mn@doi [\apjs] {10.1086/520512}, \href
  {https://ui.adsabs.harvard.edu/abs/2007ApJS..173..682M} {173, 682}

\bibitem[\protect\citeauthoryear{{Mosenkov} et~al.,}{{Mosenkov}
  et~al.}{2019}]{2019A&A...622A.132M}
{Mosenkov} A.~V.,  et~al., 2019, \mn@doi [\aap] {10.1051/0004-6361/201833932},
  \href {https://ui.adsabs.harvard.edu/abs/2019A&A...622A.132M} {622, A132}

\bibitem[\protect\citeauthoryear{{Mosenkov}, {Savchenko}, {Smirnov}  \&
  {Camps}}{{Mosenkov} et~al.}{2021}]{2021MNRAS.507.5246M}
{Mosenkov} A.~V.,  {Savchenko} S.~S.,  {Smirnov} A.~A.,   {Camps} P.,  2021,
  \mn@doi [\mnras] {10.1093/mnras/stab2445}, \href
  {https://ui.adsabs.harvard.edu/abs/2021MNRAS.507.5246M} {507, 5246}

\bibitem[\protect\citeauthoryear{{Nersesian} et~al.,}{{Nersesian}
  et~al.}{2019}]{2019A&A...624A..80N}
{Nersesian} A.,  et~al., 2019, \mn@doi [\aap] {10.1051/0004-6361/201935118},
  \href {https://ui.adsabs.harvard.edu/abs/2019A&A...624A..80N} {624, A80}

\bibitem[\protect\citeauthoryear{{Nersesian} et~al.,}{{Nersesian}
  et~al.}{2020}]{radiativeM51}
{Nersesian} A.,  et~al., 2020, \mn@doi [\aap] {10.1051/0004-6361/202038939},
  \href {https://ui.adsabs.harvard.edu/abs/2020A&A...643A..90N} {643, A90}

\bibitem[\protect\citeauthoryear{{Nersesian} et~al.,}{{Nersesian}
  et~al.}{2021}]{2021MNRAS.506.3986N}
{Nersesian} A.,  et~al., 2021, \mn@doi [\mnras] {10.1093/mnras/stab1984}, \href
  {https://ui.adsabs.harvard.edu/abs/2021MNRAS.506.3986N} {506, 3986}

\bibitem[\protect\citeauthoryear{{Noordermeer} \& {van der
  Hulst}}{{Noordermeer} \& {van der Hulst}}{2007}]{2007MNRAS.376.1480N}
{Noordermeer} E.,  {van der Hulst} J.~M.,  2007, \mn@doi [\mnras]
  {10.1111/j.1365-2966.2007.11532.x}, \href
  {https://ui.adsabs.harvard.edu/abs/2007MNRAS.376.1480N} {376, 1480}

\bibitem[\protect\citeauthoryear{{Oey}, {Parker}, {Mikles}  \& {Zhang}}{{Oey}
  et~al.}{2003}]{2003AJ....126.2317O}
{Oey} M.~S.,  {Parker} J.~S.,  {Mikles} V.~J.,   {Zhang} X.,  2003, \mn@doi
  [\aj] {10.1086/378163}, \href
  {https://ui.adsabs.harvard.edu/abs/2003AJ....126.2317O} {126, 2317}

\bibitem[\protect\citeauthoryear{{Papaderos}, {Breda}, {Humphrey}, {Michel
  Gomes}, {Ziegler}  \& {Pappalardo}}{{Papaderos}
  et~al.}{2022}]{2022A&A...658A..74P}
{Papaderos} P.,  {Breda} I.,  {Humphrey} A.,  {Michel Gomes} J.,  {Ziegler}
  B.~L.,   {Pappalardo} C.,  2022, \mn@doi [\aap]
  {10.1051/0004-6361/202140641}, \href
  {https://ui.adsabs.harvard.edu/abs/2022A&A...658A..74P} {658, A74}

\bibitem[\protect\citeauthoryear{{Peng}, {Ho}, {Impey}  \& {Rix}}{{Peng}
  et~al.}{2010}]{2010AJ....139.2097P}
{Peng} C.~Y.,  {Ho} L.~C.,  {Impey} C.~D.,   {Rix} H.-W.,  2010, \mn@doi [\aj]
  {10.1088/0004-6256/139/6/2097}, \href
  {https://ui.adsabs.harvard.edu/abs/2010AJ....139.2097P} {139, 2097}

\bibitem[\protect\citeauthoryear{{Pessa} et~al.,}{{Pessa}
  et~al.}{2023}]{2023A&A...673A.147P}
{Pessa} I.,  et~al., 2023, \mn@doi [\aap] {10.1051/0004-6361/202245673}, \href
  {https://ui.adsabs.harvard.edu/abs/2023A&A...673A.147P} {673, A147}

\bibitem[\protect\citeauthoryear{{Pilbratt} et~al.,}{{Pilbratt}
  et~al.}{2010}]{2010A&A...518L...1P}
{Pilbratt} G.~L.,  et~al., 2010, \mn@doi [\aap] {10.1051/0004-6361/201014759},
  \href {https://ui.adsabs.harvard.edu/abs/2010A&A...518L...1P} {518, L1}

\bibitem[\protect\citeauthoryear{{Popescu}, {Tuffs}, {Dopita}, {Fischera},
  {Kylafis}  \& {Madore}}{{Popescu} et~al.}{2011}]{2011A&A...527A.109P}
{Popescu} C.~C.,  {Tuffs} R.~J.,  {Dopita} M.~A.,  {Fischera} J.,  {Kylafis}
  N.~D.,   {Madore} B.~F.,  2011, \mn@doi [\aap] {10.1051/0004-6361/201015217},
  \href {https://ui.adsabs.harvard.edu/abs/2011A&A...527A.109P} {527, A109}

\bibitem[\protect\citeauthoryear{{Pour-Imani}, {Kennefick}, {Kennefick},
  {Davis}, {Shields}  \& {Shameer Abdeen}}{{Pour-Imani}
  et~al.}{2016}]{2016ApJ...827L...2P}
{Pour-Imani} H.,  {Kennefick} D.,  {Kennefick} J.,  {Davis} B.~L.,  {Shields}
  D.~W.,   {Shameer Abdeen} M.,  2016, \mn@doi [\apjl]
  {10.3847/2041-8205/827/1/L2}, \href
  {https://ui.adsabs.harvard.edu/abs/2016ApJ...827L...2P} {827, L2}

\bibitem[\protect\citeauthoryear{{Prieto}, {Beckman}, {Cepa}  \&
  {Varela}}{{Prieto} et~al.}{1992}]{1992A&A...257...85P}
{Prieto} M.,  {Beckman} J.~E.,  {Cepa} J.,   {Varela} A.~M.,  1992, \aap, \href
  {https://ui.adsabs.harvard.edu/abs/1992A&A...257...85P} {257, 85}

\bibitem[\protect\citeauthoryear{{Psychogyios} et~al.,}{{Psychogyios}
  et~al.}{2020}]{2020A&A...633A.104P}
{Psychogyios} A.,  et~al., 2020, \mn@doi [\aap] {10.1051/0004-6361/201833522},
  \href {https://ui.adsabs.harvard.edu/abs/2020A&A...633A.104P} {633, A104}

\bibitem[\protect\citeauthoryear{{Puerari}, {Elmegreen}  \& {Block}}{{Puerari}
  et~al.}{2014}]{2014AJ....148..133P}
{Puerari} I.,  {Elmegreen} B.~G.,   {Block} D.~L.,  2014, \mn@doi [\aj]
  {10.1088/0004-6256/148/6/133}, \href
  {https://ui.adsabs.harvard.edu/abs/2014AJ....148..133P} {148, 133}

\bibitem[\protect\citeauthoryear{{Querejeta} et~al.,}{{Querejeta}
  et~al.}{2016}]{2016A&A...593A.118Q}
{Querejeta} M.,  et~al., 2016, \mn@doi [\aap] {10.1051/0004-6361/201628674},
  \href {https://ui.adsabs.harvard.edu/abs/2016A&A...593A.118Q} {593, A118}

\bibitem[\protect\citeauthoryear{{R{\'\i}os-L{\'o}pez}, {A{\~n}orve},
  {Ibarra-Medel}, {L{\'o}pez-Cruz}, {Alvira-Enr{\'\i}quez}, {Iacobuta}  \&
  {Valerdi}}{{R{\'\i}os-L{\'o}pez} et~al.}{2021}]{2021MNRAS.507.5952R}
{R{\'\i}os-L{\'o}pez} E.,  {A{\~n}orve} C.,  {Ibarra-Medel} H.~J.,
  {L{\'o}pez-Cruz} O.,  {Alvira-Enr{\'\i}quez} J.,  {Iacobuta} G.,   {Valerdi}
  M.,  2021, \mn@doi [\mnras] {10.1093/mnras/stab2321}, \href
  {https://ui.adsabs.harvard.edu/abs/2021MNRAS.507.5952R} {507, 5952}

\bibitem[\protect\citeauthoryear{{Robotham}, {Taranu}, {Tobar}, {Moffett}  \&
  {Driver}}{{Robotham} et~al.}{2017}]{2017MNRAS.466.1513R}
{Robotham} A.~S.~G.,  {Taranu} D.~S.,  {Tobar} R.,  {Moffett} A.,   {Driver}
  S.~P.,  2017, \mn@doi [\mnras] {10.1093/mnras/stw3039}, \href
  {https://ui.adsabs.harvard.edu/abs/2017MNRAS.466.1513R} {466, 1513}

\bibitem[\protect\citeauthoryear{{Robotham}, {Bellstedt}  \&
  {Driver}}{{Robotham} et~al.}{2022}]{2022MNRAS.513.2985R}
{Robotham} A.~S.~G.,  {Bellstedt} S.,   {Driver} S.~P.,  2022, \mn@doi [\mnras]
  {10.1093/mnras/stac1032}, \href
  {https://ui.adsabs.harvard.edu/abs/2022MNRAS.513.2985R} {513, 2985}

\bibitem[\protect\citeauthoryear{{Rosse}}{{Rosse}}{1850}]{1850RSPT..140..499R}
{Rosse} T. E.~O.,  1850, Philosophical Transactions of the Royal Society of
  London Series I, \href
  {https://ui.adsabs.harvard.edu/abs/1850RSPT..140..499R} {140, 499}

\bibitem[\protect\citeauthoryear{{Salo} \& {Laurikainen}}{{Salo} \&
  {Laurikainen}}{2000}]{2000MNRAS.319..377S}
{Salo} H.,  {Laurikainen} E.,  2000, \mn@doi [\mnras]
  {10.1046/j.1365-8711.2000.03650.x}, \href
  {https://ui.adsabs.harvard.edu/abs/2000MNRAS.319..377S} {319, 377}

\bibitem[\protect\citeauthoryear{{Salo} et~al.,}{{Salo}
  et~al.}{2015}]{2015ApJS..219....4S}
{Salo} H.,  et~al., 2015, \mn@doi [\apjs] {10.1088/0067-0049/219/1/4}, \href
  {https://ui.adsabs.harvard.edu/abs/2015ApJS..219....4S} {219, 4}

\bibitem[\protect\citeauthoryear{{Savchenko} \& {Reshetnikov}}{{Savchenko} \&
  {Reshetnikov}}{2013}]{2013MNRAS.436.1074S}
{Savchenko} S.~S.,  {Reshetnikov} V.~P.,  2013, \mn@doi [\mnras]
  {10.1093/mnras/stt1627}, \href
  {https://ui.adsabs.harvard.edu/abs/2013MNRAS.436.1074S} {436, 1074}

\bibitem[\protect\citeauthoryear{{Savchenko}, {Marchuk}, {Mosenkov}  \&
  {Grishunin}}{{Savchenko} et~al.}{2020}]{savchenko}
{Savchenko} S.,  {Marchuk} A.,  {Mosenkov} A.,   {Grishunin} K.,  2020, \mn@doi
  [\mnras] {10.1093/mnras/staa258}, \href
  {https://ui.adsabs.harvard.edu/abs/2020MNRAS.493..390S} {493, 390}

\bibitem[\protect\citeauthoryear{{Scarano} \& {L{\'e}pine}}{{Scarano} \&
  {L{\'e}pine}}{2013}]{2013MNRAS.428..625S}
{Scarano} S.,  {L{\'e}pine} J.~R.~D.,  2013, \mn@doi [\mnras]
  {10.1093/mnras/sts048}, \href
  {https://ui.adsabs.harvard.edu/abs/2013MNRAS.428..625S} {428, 625}

\bibitem[\protect\citeauthoryear{{Scheepmaker}, {Lamers}, {Anders}  \&
  {Larsen}}{{Scheepmaker} et~al.}{2009}]{2009A&A...494...81S}
{Scheepmaker} R.~A.,  {Lamers} H.~J.~G.~L.~M.,  {Anders} P.,   {Larsen} S.~S.,
  2009, \mn@doi [\aap] {10.1051/0004-6361:200811068}, \href
  {https://ui.adsabs.harvard.edu/abs/2009A&A...494...81S} {494, 81}

\bibitem[\protect\citeauthoryear{{Schwarz}}{{Schwarz}}{1978}]{1978AnSta...6..461S}
{Schwarz} G.,  1978, Annals of Statistics, \href
  {https://ui.adsabs.harvard.edu/abs/1978AnSta...6..461S} {6, 461}

\bibitem[\protect\citeauthoryear{{Sersic}}{{Sersic}}{1968}]{1968adga.book.....S}
{Sersic} J.~L.,  1968, {Atlas de Galaxias Australes}

\bibitem[\protect\citeauthoryear{{Shabani} et~al.,}{{Shabani}
  et~al.}{2018}]{2018MNRAS.478.3590S}
{Shabani} F.,  et~al., 2018, \mn@doi [\mnras] {10.1093/mnras/sty1277}, \href
  {https://ui.adsabs.harvard.edu/abs/2018MNRAS.478.3590S} {478, 3590}

\bibitem[\protect\citeauthoryear{{Sheth} et~al.,}{{Sheth}
  et~al.}{2008}]{2008ApJ...675.1141S}
{Sheth} K.,  et~al., 2008, \mn@doi [\apj] {10.1086/524980}, \href
  {https://ui.adsabs.harvard.edu/abs/2008ApJ...675.1141S} {675, 1141}

\bibitem[\protect\citeauthoryear{{Sheth} et~al.,}{{Sheth}
  et~al.}{2010}]{2010PASP..122.1397S}
{Sheth} K.,  et~al., 2010, \mn@doi [\pasp] {10.1086/657638}, \href
  {https://ui.adsabs.harvard.edu/abs/2010PASP..122.1397S} {122, 1397}

\bibitem[\protect\citeauthoryear{{Shetty}, {Vogel}, {Ostriker}  \&
  {Teuben}}{{Shetty} et~al.}{2007}]{2007ApJ...665.1138S}
{Shetty} R.,  {Vogel} S.~N.,  {Ostriker} E.~C.,   {Teuben} P.~J.,  2007,
  \mn@doi [\apj] {10.1086/520037}, \href
  {https://ui.adsabs.harvard.edu/abs/2007ApJ...665.1138S} {665, 1138}

\bibitem[\protect\citeauthoryear{{Simard}, {Mendel}, {Patton}, {Ellison}  \&
  {McConnachie}}{{Simard} et~al.}{2011}]{2011ApJS..196...11S}
{Simard} L.,  {Mendel} J.~T.,  {Patton} D.~R.,  {Ellison} S.~L.,
  {McConnachie} A.~W.,  2011, \mn@doi [\apjs] {10.1088/0067-0049/196/1/11},
  \href {https://ui.adsabs.harvard.edu/abs/2011ApJS..196...11S} {196, 11}

\bibitem[\protect\citeauthoryear{{Skrutskie} et~al.,}{{Skrutskie}
  et~al.}{2006}]{2006AJ....131.1163S}
{Skrutskie} M.~F.,  et~al., 2006, \mn@doi [\aj] {10.1086/498708}, \href
  {https://ui.adsabs.harvard.edu/abs/2006AJ....131.1163S} {131, 1163}

\bibitem[\protect\citeauthoryear{{Smirnov} \& {Savchenko}}{{Smirnov} \&
  {Savchenko}}{2020}]{2020MNRAS.499..462S}
{Smirnov} A.~A.,  {Savchenko} S.~S.,  2020, \mn@doi [\mnras]
  {10.1093/mnras/staa2892}, \href
  {https://ui.adsabs.harvard.edu/abs/2020MNRAS.499..462S} {499, 462}

\bibitem[\protect\citeauthoryear{{Sofue}, {Tutui}, {Honma}, {Tomita},
  {Takamiya}, {Koda}  \& {Takeda}}{{Sofue} et~al.}{1999}]{1999ApJ...523..136S}
{Sofue} Y.,  {Tutui} Y.,  {Honma} M.,  {Tomita} A.,  {Takamiya} T.,  {Koda} J.,
    {Takeda} Y.,  1999, \mn@doi [\apj] {10.1086/307731}, \href
  {https://ui.adsabs.harvard.edu/abs/1999ApJ...523..136S} {523, 136}

\bibitem[\protect\citeauthoryear{{Somerville} \& {Dav{\'e}}}{{Somerville} \&
  {Dav{\'e}}}{2015}]{2015ARA&A..53...51S}
{Somerville} R.~S.,  {Dav{\'e}} R.,  2015, \mn@doi [\araa]
  {10.1146/annurev-astro-082812-140951}, \href
  {https://ui.adsabs.harvard.edu/abs/2015ARA&A..53...51S} {53, 51}

\bibitem[\protect\citeauthoryear{{Sonnenfeld}}{{Sonnenfeld}}{2022}]{2022A&A...659A.141S}
{Sonnenfeld} A.,  2022, \mn@doi [\aap] {10.1051/0004-6361/202142786}, \href
  {https://ui.adsabs.harvard.edu/abs/2022A&A...659A.141S} {659, A141}

\bibitem[\protect\citeauthoryear{{Tamburro}, {Rix}, {Walter}, {Brinks}, {de
  Blok}, {Kennicutt}  \& {Mac Low}}{{Tamburro} et~al.}{2008}]{Tamburro2008}
{Tamburro} D.,  {Rix} H.~W.,  {Walter} F.,  {Brinks} E.,  {de Blok} W.~J.~G.,
  {Kennicutt} R.~C.,   {Mac Low} M.~M.,  2008, \mn@doi [\aj]
  {10.1088/0004-6256/136/6/2872}, \href
  {https://ui.adsabs.harvard.edu/abs/2008AJ....136.2872T} {136, 2872}

\bibitem[\protect\citeauthoryear{{Terashima} \& {Wilson}}{{Terashima} \&
  {Wilson}}{2001}]{2001ApJ...560..139T}
{Terashima} Y.,  {Wilson} A.~S.,  2001, \mn@doi [\apj] {10.1086/321615}, \href
  {https://ui.adsabs.harvard.edu/abs/2001ApJ...560..139T} {560, 139}

\bibitem[\protect\citeauthoryear{{Toomre} \& {Toomre}}{{Toomre} \&
  {Toomre}}{1972}]{1972ApJ...178..623T}
{Toomre} A.,  {Toomre} J.,  1972, \mn@doi [\apj] {10.1086/151823}, \href
  {https://ui.adsabs.harvard.edu/abs/1972ApJ...178..623T} {178, 623}

\bibitem[\protect\citeauthoryear{{Trujillo}, {Aguerri}, {Cepa}  \&
  {Guti{\'e}rrez}}{{Trujillo} et~al.}{2001a}]{2001MNRAS.321..269T}
{Trujillo} I.,  {Aguerri} J.~A.~L.,  {Cepa} J.,   {Guti{\'e}rrez} C.~M.,
  2001a, \mn@doi [\mnras] {10.1046/j.1365-8711.2001.03987.x}, \href
  {https://ui.adsabs.harvard.edu/abs/2001MNRAS.321..269T} {321, 269}

\bibitem[\protect\citeauthoryear{{Trujillo}, {Aguerri}, {Cepa}  \&
  {Guti{\'e}rrez}}{{Trujillo} et~al.}{2001b}]{2001MNRAS.328..977T}
{Trujillo} I.,  {Aguerri} J.~A.~L.,  {Cepa} J.,   {Guti{\'e}rrez} C.~M.,
  2001b, \mn@doi [\mnras] {10.1046/j.1365-8711.2001.04937.x}, \href
  {https://ui.adsabs.harvard.edu/abs/2001MNRAS.328..977T} {328, 977}

\bibitem[\protect\citeauthoryear{{Tr{\v{c}}ka} et~al.,}{{Tr{\v{c}}ka}
  et~al.}{2020}]{2020MNRAS.494.2823T}
{Tr{\v{c}}ka} A.,  et~al., 2020, \mn@doi [\mnras] {10.1093/mnras/staa857},
  \href {https://ui.adsabs.harvard.edu/abs/2020MNRAS.494.2823T} {494, 2823}

\bibitem[\protect\citeauthoryear{{Tully}}{{Tully}}{1974}]{1974ApJS...27..449T}
{Tully} R.~B.,  1974, \mn@doi [\apjs] {10.1086/190306}, \href
  {https://ui.adsabs.harvard.edu/abs/1974ApJS...27..449T} {27, 449}

\bibitem[\protect\citeauthoryear{{Vall{\'e}e}}{{Vall{\'e}e}}{2020}]{2020NewA...7601337V}
{Vall{\'e}e} J.~P.,  2020, \mn@doi [\na] {10.1016/j.newast.2019.101337}, \href
  {https://ui.adsabs.harvard.edu/abs/2020NewA...7601337V} {76, 101337}

\bibitem[\protect\citeauthoryear{{Verstocken} et~al.,}{{Verstocken}
  et~al.}{2020}]{radiativeM81}
{Verstocken} S.,  et~al., 2020, \mn@doi [\aap] {10.1051/0004-6361/201935770},
  \href {https://ui.adsabs.harvard.edu/abs/2020A&A...637A..24V} {637, A24}

\bibitem[\protect\citeauthoryear{{Vika}, {Driver}, {Cameron}, {Kelvin}  \&
  {Robotham}}{{Vika} et~al.}{2012}]{2012MNRAS.419.2264V}
{Vika} M.,  {Driver} S.~P.,  {Cameron} E.,  {Kelvin} L.,   {Robotham} A.,
  2012, \mn@doi [\mnras] {10.1111/j.1365-2966.2011.19881.x}, \href
  {https://ui.adsabs.harvard.edu/abs/2012MNRAS.419.2264V} {419, 2264}

\bibitem[\protect\citeauthoryear{{Vika}, {Bamford}, {H{\"a}u{\ss}ler}  \&
  {Rojas}}{{Vika} et~al.}{2014}]{2014MNRAS.444.3603V}
{Vika} M.,  {Bamford} S.~P.,  {H{\"a}u{\ss}ler} B.,   {Rojas} A.~L.,  2014,
  \mn@doi [\mnras] {10.1093/mnras/stu1696}, \href
  {https://ui.adsabs.harvard.edu/abs/2014MNRAS.444.3603V} {444, 3603}

\bibitem[\protect\citeauthoryear{{Vlahakis}, {van der Werf}, {Israel}  \&
  {Tilanus}}{{Vlahakis} et~al.}{2013}]{2013MNRAS.433.1837V}
{Vlahakis} C.,  {van der Werf} P.,  {Israel} F.~P.,   {Tilanus} R.~P.~J.,
  2013, \mn@doi [\mnras] {10.1093/mnras/stt841}, \href
  {https://ui.adsabs.harvard.edu/abs/2013MNRAS.433.1837V} {433, 1837}

\bibitem[\protect\citeauthoryear{{Vollmer} \& {Leroy}}{{Vollmer} \&
  {Leroy}}{2011}]{2011AJ....141...24V}
{Vollmer} B.,  {Leroy} A.~K.,  2011, \mn@doi [\aj]
  {10.1088/0004-6256/141/1/24}, \href
  {https://ui.adsabs.harvard.edu/abs/2011AJ....141...24V} {141, 24}

\bibitem[\protect\citeauthoryear{{Vulcani} et~al.,}{{Vulcani}
  et~al.}{2014}]{2014MNRAS.441.1340V}
{Vulcani} B.,  et~al., 2014, \mn@doi [\mnras] {10.1093/mnras/stu632}, \href
  {https://ui.adsabs.harvard.edu/abs/2014MNRAS.441.1340V} {441, 1340}

\bibitem[\protect\citeauthoryear{{Walmsley} et~al.,}{{Walmsley}
  et~al.}{2023}]{2023MNRAS.526.4768W}
{Walmsley} M.,  et~al., 2023, \mn@doi [\mnras] {10.1093/mnras/stad2919}, \href
  {https://ui.adsabs.harvard.edu/abs/2023MNRAS.526.4768W} {526, 4768}

\bibitem[\protect\citeauthoryear{{Watkins}, {Laine}, {Comer{\'o}n}, {Janz}  \&
  {Salo}}{{Watkins} et~al.}{2019}]{2019A&A...625A..36W}
{Watkins} A.~E.,  {Laine} J.,  {Comer{\'o}n} S.,  {Janz} J.,   {Salo} H.,
  2019, \mn@doi [\aap] {10.1051/0004-6361/201935130}, \href
  {https://ui.adsabs.harvard.edu/abs/2019A&A...625A..36W} {625, A36}

\bibitem[\protect\citeauthoryear{{Weinzirl}, {Jogee}, {Khochfar}, {Burkert}  \&
  {Kormendy}}{{Weinzirl} et~al.}{2009}]{2009ApJ...696..411W}
{Weinzirl} T.,  {Jogee} S.,  {Khochfar} S.,  {Burkert} A.,   {Kormendy} J.,
  2009, \mn@doi [\apj] {10.1088/0004-637X/696/1/411}, \href
  {https://ui.adsabs.harvard.edu/abs/2009ApJ...696..411W} {696, 411}

\bibitem[\protect\citeauthoryear{{Werner} et~al.,}{{Werner}
  et~al.}{2004}]{2004ApJS..154....1W}
{Werner} M.~W.,  et~al., 2004, \mn@doi [\apjs] {10.1086/422992}, \href
  {https://ui.adsabs.harvard.edu/abs/2004ApJS..154....1W} {154, 1}

\bibitem[\protect\citeauthoryear{{Willett} et~al.,}{{Willett}
  et~al.}{2013}]{2013MNRAS.435.2835W}
{Willett} K.~W.,  et~al., 2013, \mn@doi [\mnras] {10.1093/mnras/stt1458}, \href
  {https://ui.adsabs.harvard.edu/abs/2013MNRAS.435.2835W} {435, 2835}

\bibitem[\protect\citeauthoryear{{Williams} \& {Evans}}{{Williams} \&
  {Evans}}{2017}]{2017MNRAS.469.4414W}
{Williams} A.~A.,  {Evans} N.~W.,  2017, \mn@doi [\mnras]
  {10.1093/mnras/stx1198}, \href
  {https://ui.adsabs.harvard.edu/abs/2017MNRAS.469.4414W} {469, 4414}

\bibitem[\protect\citeauthoryear{{Wright} et~al.,}{{Wright}
  et~al.}{2010}]{2010AJ....140.1868W}
{Wright} E.~L.,  et~al., 2010, \mn@doi [\aj] {10.1088/0004-6256/140/6/1868},
  \href {https://ui.adsabs.harvard.edu/abs/2010AJ....140.1868W} {140, 1868}

\bibitem[\protect\citeauthoryear{{York} et~al.,}{{York}
  et~al.}{2000}]{2000AJ....120.1579Y}
{York} D.~G.,  et~al., 2000, \mn@doi [\aj] {10.1086/301513}, \href
  {https://ui.adsabs.harvard.edu/abs/2000AJ....120.1579Y} {120, 1579}

\bibitem[\protect\citeauthoryear{{Yu} \& {Ho}}{{Yu} \&
  {Ho}}{2018}]{2018ApJ...869...29Y}
{Yu} S.-Y.,  {Ho} L.~C.,  2018, \mn@doi [\apj] {10.3847/1538-4357/aaeacd},
  \href {https://ui.adsabs.harvard.edu/abs/2018ApJ...869...29Y} {869, 29}

\bibitem[\protect\citeauthoryear{{Yu}, {Ho}, {Barth}  \& {Li}}{{Yu}
  et~al.}{2018}]{2018ApJ...862...13Y}
{Yu} S.-Y.,  {Ho} L.~C.,  {Barth} A.~J.,   {Li} Z.-Y.,  2018, \mn@doi [\apj]
  {10.3847/1538-4357/aacb25}, \href
  {https://ui.adsabs.harvard.edu/abs/2018ApJ...862...13Y} {862, 13}

\bibitem[\protect\citeauthoryear{{de Vaucouleurs}, {de Vaucouleurs}, {Corwin},
  {Buta}, {Paturel}  \& {Fouque}}{{de Vaucouleurs}
  et~al.}{1995}]{1995yCat.7155....0D}
{de Vaucouleurs} G.,  {de Vaucouleurs} A.,  {Corwin} H.~G.,  {Buta} R.~J.,
  {Paturel} G.,   {Fouque} P.,  1995, VizieR Online Data Catalog, \href
  {https://ui.adsabs.harvard.edu/abs/1995yCat.7155....0D} {p. VII/155}

\makeatother
\end{thebibliography}

\section*{Appendix A}
\label{appendix}

Available in the online supplement material. Here, we present Tables for parameters of traditional components in remaining datasets for both sets of models B+D and B+D+S. We also list parameters of spiral arms, obtained for those images.

\bsp	
\label{lastpage}
\end{document}